\documentclass[useAMS,usenatbib]{mn2e}
\usepackage{graphicx}

\title[A photometric study of the southern Blazhko star SS For]{A photometric
  study of the southern Blazhko star SS For\\
  Unambiguous detection of quintuplet components}
\author[K. Kolenberg, et al.]
{K. Kolenberg$^{1}$\thanks{E-mail:
kolenberg@astro.univie.ac.at}
E. Guggenberger$^{1}$,
T. Medupe$^{2}$,
P. Lenz$^{1}$,
L. Schmitzberger$^{1}$,
\and
R.R. Shobbrook$^{3}$,
P. Beck$^{1}$,
B. Ngwato$^{4}$,
and
J. Lub$^{5}$,
\\
$^{1}$Institute of Astronomy, T\"urkenschanzstrasse 17, A-1180 Vienna, 
Austria\\
$^{2}$Astronomy Department, University of Cape Town, South Africa\\
$^{3}$Research School of Astronomy and Astrophysics, Australian National University, Canberra, ACT, Australia\\
$^{4}$Theoretical Astrophysics Programme, North West University, Mmabatho,
 South Africa\\
 $^{5}$Leiden Observatory, Niels Bohrweg 2, NL-2333 CA Leiden, The Netherlands}
\begin{document}

\date{Received 2008 October 8, Accepted 2008 November 18.}

\maketitle

\begin{abstract}
We present our analysis of photometric data in the Johnson $B$ and $V$ filter of the 
southern Blazhko star SS For.
In parallel, we analyzed the $V$ observations obtained with the
ASAS-3 photometry of the star gathered between 2000 and 2008.  
In the frequency spectra resulting from a Fourier analysis of our data, the triplet structure is detectable up to high order,
both in the $B$ and $V$ data.  
Moreover, we find evidence for quintuplet components.   
We confirm from our data that the modulation
components decrease less steeply than the harmonics of the main frequency. 
We derived the
variations of the Fourier parameters quantifying the light curve shape over the Blazhko cycle.
There is good agreement between the spectroscopic abundance and
the metallicity determined from
the Fourier parameters of the average light curve.
SS For is peculiar as a Blazhko star because of its strong variations around
minimum light. 
\end{abstract}


\section{Introduction}
A large fraction of the RR Lyrae stars shows a periodic amplitude and/or phase
modulation with a period of typically ten to hundreds of times the pulsation
period. This phenomenon is referred to as the Blazhko effect, after the
Russian astronomer who first reported it (Blazhko 1907).  The origin of the Blazhko effect is still a
matter of controversy.  The most widely discussed models attribute the effect 
to either the consequences of the interaction of a magnetic field with the 
main radial pulsation (Shibahashi \& Takata 1995), or a resonance between the 
main mode and a non-radial mode of low degree (Van Hoolst, Dziembowski \& Kawaler 1998;
Nowakowski \& Dziembowski 2001; Dziembowski \& Mizerski 2004). However, also models that do not require nonradial modes have been proposed, e.g. by Stothers (2006). This scenario attributes the Blazhko variation to a variable turbulent convection due to transient magnetic fields in the star.
All models presently proposed for the Blazhko effect have shortcomings in explaining the variety of features shown by Blazhko stars.  Therefore, theoretical efforts to revise or expand the models would be worthwhile, and even the exploration of alternative explanations. 

Since its discovery almost a century ago, most studies of the Blazhko effect 
were carried out from the northern hemisphere (e.g., Szeidl 1988, and
references therein; Smith 1995, and references therein). 
For a long time there were fewer well-established
{\it field} Blazhko stars at southern declinations.  The available data for
southern field Blazhko stars (e.g., Hoffmeister 1956; Kinman 1961; Clube et
al. 1969; Lub 1977) are often insufficient to determine the Blazhko 
periods with the required accuracy.
Accurate and
complete photometric data sets of southern field Blazhko stars are
still lacking.
Extensive multi-target surveys such as MACHO (Alcock et al. 
2000, 2003) and OGLE (Moskalik \& Poretti 2003), and ASAS (All Sky Automated
Survey, Pojmanski 2000)
have significantly contributed to our knowledge of the Blazhko effect, and
expanded the list of known Blazhko stars.  
Wils \& Sodor (2005) published a list of new 
and confirmed Blazhko targets, and Szczygiel \& Fabrycky (2007) obtained interesting new results, among which is a star with multiple Blazhko periods. 

Nevertheless, long-term campaigns dedicated to particular Blazhko stars and
yielding complete light curves at different phases in the Blazhko cycle
remain of great scientific value.  Only these can reveal changes in the light
curve shape, as well as long-term changes 
in the characteristics of the Blazhko effect such as amplitudes, phases, and
periods. In this way they can give crucial information for deciding among the
different hypotheses for the Blazhko effect.  

Photometry with a good spread over both the pulsation and
Blazhko cycles allows us to perform the first rigorous frequency 
analysis of a southern field Blazhko star.  
  The first accurate photometric data
set covering both the pulsation and the Blazhko cycle was published by Jurcsik
et al. (2005a) for the northern short-period Blazhko star RR Gem.  

SS For (SAO 167572; $\alpha(J2000)$: 02$^{\rm h}$ 07$^{\rm m}$ 52$^{\rm
  s}$, $\delta(J2000)$: -26$^{\circ}$ 51' 54'') is a relatively bright 
southern Blazhko star. According to the General Catalogue of Variable Stars
  (Kukarkin 2003) 
its $V$ brightness changes within the range 9.45-10.60.
The General Catalogue of Variable Stars (GCVS) lists a period of $P$ = 0.495432 d (Kholopov et al. 1998) or
about 11 h 53 min, 
corresponding to a frequency of $f_0 = 2.0184405$ cd$^{\rm -1}$.

We selected this target on the basis of references to its variable
light curve in the literature (Lub 1977). Our analysis of both the HIPPARCOS and the ASAS-3 
photometry demonstrated the existence of a Blazhko effect with strong
variations around minimum.  The deduced Blazhko period of SS For 
(about 35 d) turned out to be short enough to cover several modulation cycles during one observing season.
All our data of SS For are available online at http://www.univie.ac.at/tops/blazhko/data/.

In Section\,2 we present the new observations as
well as the ASAS-3 data used for comparison.  The different steps in the 
data analysis and its results are described in Section\,3.
Section\,4 is dedicated to a discussion of the results obtained. Finally, some
concluding remarks are given in Section\,5.

\section{Observations}

\subsection{New photometric measurements}

Our photometric observations were carried out with 3 different telescopes in 
the southern hemisphere.
All were equipped with photomultipliers to obtain comparative photometry in
the Johnson $B$ and $V$ passbands.

The photometric observations of SS For started from the South African Astronomical Observatory
(SAAO) in Sutherland, South Africa, in October 2004, and were carried out from
the 0.5-m and 0.75-m telescopes until November 2005.  The 0.5-m telescope was equipped with the
modular photometer.
At the 0.75-m telescope we used the UCT photometer.
From July until September 2005, the 0.6-m telescope at Siding Spring Observatory (SSO) near Coonabarabran, Australia,
joined in the observing campaign. Its photometer is equipped with a GaAs
photomultiplier tube.
Given the nearly 12-h period of SS For, the complementary longitudes of the two observatories allowed for observations
of a larger part of the light curve.  Additionally, the height of 1-day
aliases introduced by the spectral window (see also Fig.\,1) was decreased by
the use of more than one observing site.
This enables an easier and more accurate determination of the 
frequencies of the stellar light variations.  The total time span of the data
set is 345 days, almost 10 Blazhko cycles.

The observations were carried out using the 3-star technique, as described by
Breger (1993). 
The comparison and check stars were HD13334 
($\alpha(J2000)$: 02$^{\rm h}$ 09$^{\rm m}$ 45$^{\rm
  s}$, $\delta(J2000)$: -27$^{\circ}$ 06' 13'',
$V$ =9.7, $B-V$ = 0.57) and HD13181 ($\alpha(J2000)$: 02$^{\rm h}$ 08$^{\rm m}$ 13$^{\rm
  s}$, $\delta(J2000)$: -27$^{\circ}$ 01' 29'', $V$ =9.8, $B-V$ = 0.51).
Table\,1 shows a journal of the measurements. 
Depending on the weather conditions, the accuracy of the individual
differential measurements varied between 3 and 10 mmag. 
In the reduction procedure, variable atmospheric extinction was taken into
account. The night-to-night shifts of the differential magnitudes of the two
comparison stars laid within the scatter of a few
millimags. 

In order to combine the data sets from the different telescopes and
instruments we carried out transformations 
to the standard system.  This is crucial, since for variables with changing
light curve shapes zero point offsets can often not be correctly determined.
Moreover, the large variations in the B-V colour (more than 0.3 mag - see
Figure\,3)
throughout the pulsation cycle result in a considerable distortion of the light curves (as large
as 0.02-0.03 mag) if the colour systems are not the same.
For the SAAO data we used transformation coefficients determined by
D. Kilkenny (private communication), and from our own standard star
measurements.
For the SSO data we took coefficients determined from standard star measurements by
R. Shobbrook (private communication), close to the ones given by Berndnikov \& Turner (2001). 

The light curves of the new data in Johnson $B$ and Johnson $V$, folded with
the main pulsation period, are shown in Figure\,2. The differential magnitudes
are shown with respect to HD13181.

\begin{figure} 
\includegraphics[width=9cm]{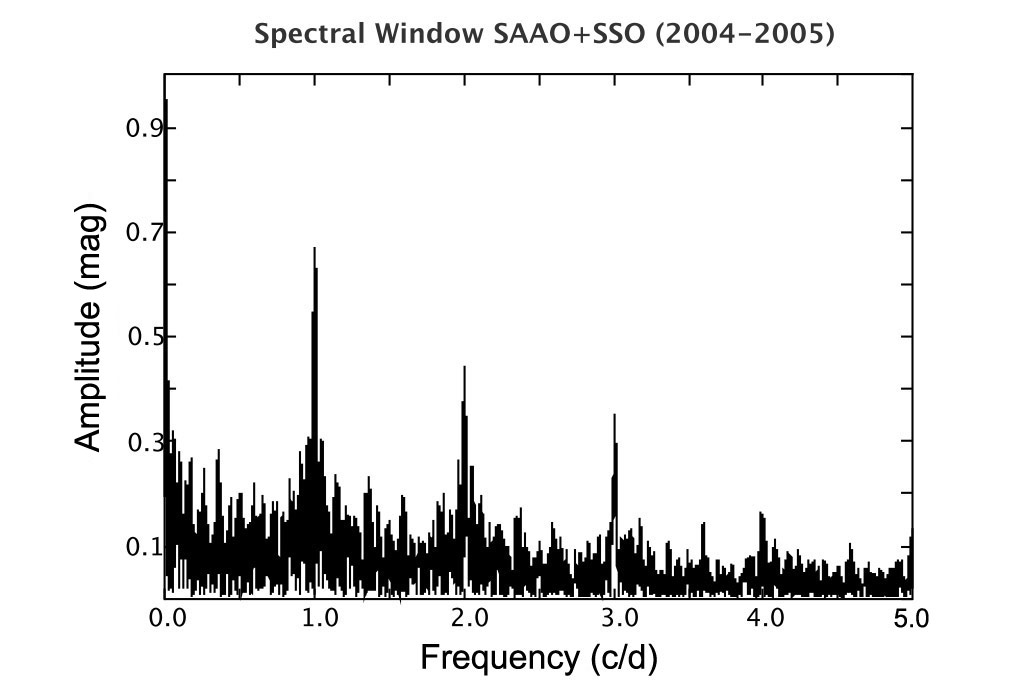}
 \caption{Window Function for the data set gathered at SAAO and SSO (2004-2005). }
\end{figure}


\begin{table*} 
\caption[]{Observing log for the SS For Johnson $B$ and $V$ data.}
\begin{center}
\begin{tabular}{cccl|cccl}
\hline
\hline
\multicolumn{4}{c|}{{\bf SAAO data} (2004+2005)} & \multicolumn{4}{c}{{\bf SSO
    data} (2005)}\\
Night &	Length [hrs] &	$N$	& Observer &	Night &	Length [hrs] & $N$ & Observer \\	
\hline
2453288&	2.73&	18&	EG&		2453569&	0.98&	6&
PL\\ 	
2453289&	7.30&	43&	EG&		2453571&	2.36&	11 &
PL\\ 
2453290&	8.57&	62&	EG&		2453574&	2.39&	10 &
PL\\ 	
2453292&	3.54&	24&	EG&		2453576&	2.30&	11 &
PL\\ 
2453296&	6.07&	44&	EG&		2453580&	2.61&	12 &
PL\\	
2453300&	5.92&	21&	EG&		2453581&	2.73&	13 &
PL\\	
2453301&	8.29&	60&	EG&		2453582&	2.59&	14 &
PL\\	
2453302&	3.67&	23&	EG&		2453583&	2.66&	14 &
PL\\	
2453303&	7.07&	44&	EG&		2453590&	2.66&	12 &
LS\\	
2453304&	7.59&	51&	EG&		2453591&	2.91&	12 &
LS\\	
2453319&	4.01&	18&	TM&		2453610&	5.76&	27 &
LS\\	
2453320&	3.74&	17&	TM&		2453611&	4.81&	22 &
LS\\	
2453321&	4.55&	19&	TM&		2453612&	1.78&	9 &
LS\\ 	
2453324&	3.75&	18&	TM&		2453621&	1.81&	10 &
LS\\	
2453325&	4.13&	22&	TM&		2453622&	1.58&	9&
LS\\ 	
2453367&	1.69&	6	&EG	&	2453633	&3.57	&26&	BS\\ 	
2453574&	4.29&	26&	EG& &&&\\
2453575&	4.12&	23&	EG& &&&\\						
2453576&	2.82&	20&	EG& &&&\\						
2453578&	0.54&	4	&EG	& &&&\\					
2453579&	3.17&	21&	EG& &&&\\						
2453580&	4.86&	29&	EG& &&&\\						
2453582&	4.81&	31&	EG& &&&\\						
2453584&	4.51&	26&	EG& &&&\\						
2453587&	2.11&	12&	PB& &&&\\						
2453588&	1.48&	7	&PB	& &&&\\					
2453592&	3.86&	26&	EG& &&&\\						
2453593&	5.18&	34&	EG& &&&\\						
2453594&	5.38&	36&	EG& &&&\\						
2453596&	5.40&	36&	EG& &&&\\						
2453598&	5.66&	40&	EG& &&&\\						
2453599&	5.60&	45&	EG& &&&\\						
2453601&	5.90&	48&	EG& &&&\\					
2453602&	6.00&	46&	EG& &&&\\
\hline
\hline
& & & &&&&\\							
{\bf SAAO} &		total&	2004&	2005&	{\bf SSO}	&
	\multicolumn{3}{c}{total (=2005)}\\	
	\vspace{0.5mm}	\\
\multicolumn{1}{l}{Observing time [hrs]}	&	158.32&	82.63&	75.70&	\multicolumn{1}{l}{Observing time [hrs]}&\multicolumn{3}{c}{43.51}\\	
\multicolumn{1}{l}{Number of datapoints}    &	1000 &	490 &	510&\multicolumn{1}{l}{	Number of datapoints}&\multicolumn{3}{c}{			218}\\	
\multicolumn{1}{l}{Number of nights}	&	34   &	16  &	18 &\multicolumn{1}{l}{	Number of nights} &\multicolumn{3}{c}{			16}	\\
\multicolumn{1}{l}{Total time span [days]}	&	314  &	79  &	28 &\multicolumn{1}{l}{	Total time span
[days]}&\multicolumn{3}{c}{64	} \\
\vspace{3mm}
\\										
&\multicolumn{3}{l}{		Total observing time [hrs]}&\multicolumn{4}{l}{				201.83	}\\		
&\multicolumn{3}{l}{		Total number of datapoints}&\multicolumn{4}{l}{				1218}\\			
&\multicolumn{3}{l}{		Number of nights}&\multicolumn{4}{l}{				50}\\			
&\multicolumn{3}{l}{		Total time span [days]}&\multicolumn{4}{l}{			345}\\			
\hline
\end{tabular}
\end{center}
\end{table*}

\begin{figure} 
\includegraphics[width=9cm]{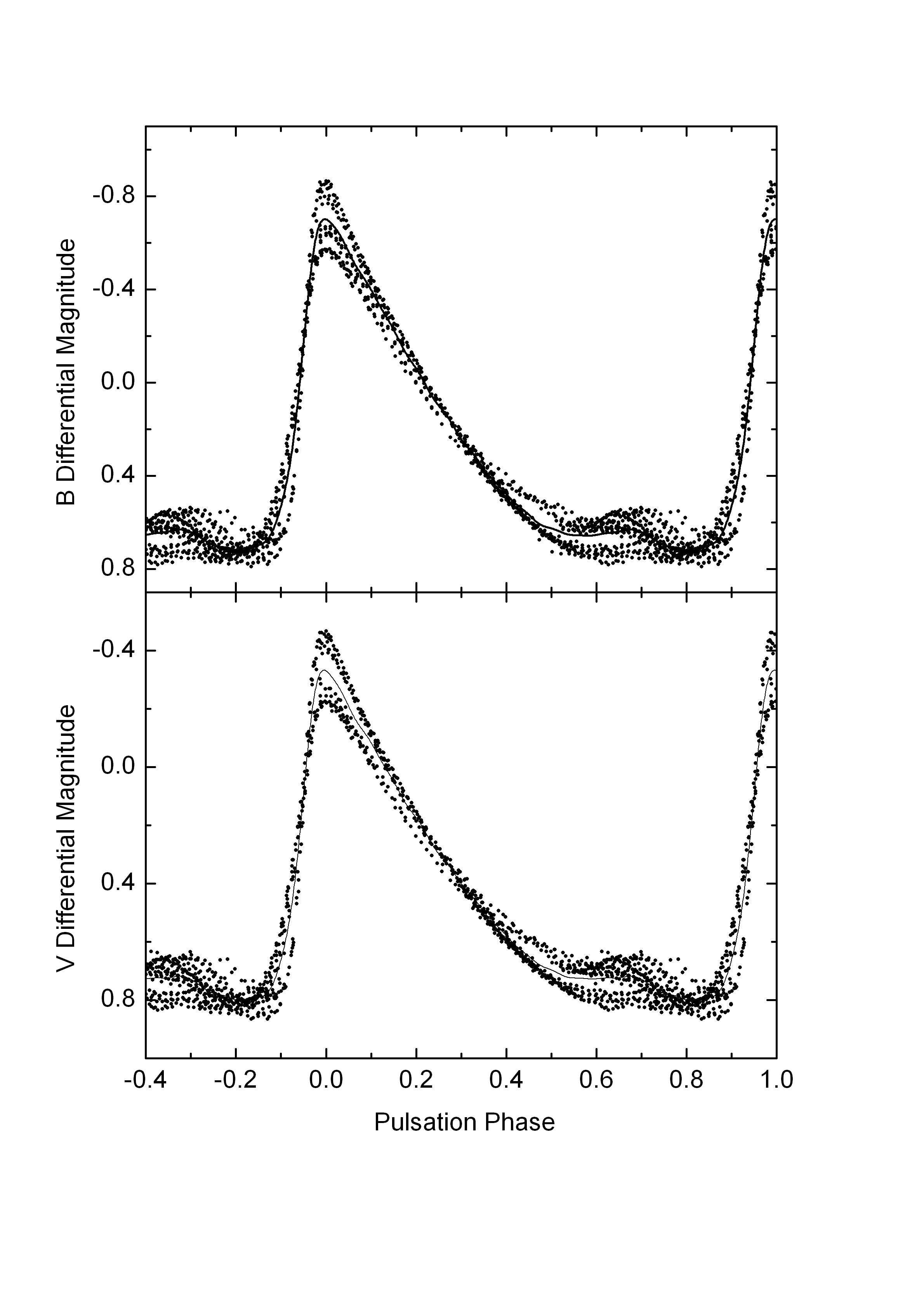}
 \caption{$B$ (upper panel) and $V$ (lower panel) light curve of SS For,
 folded with the main period.  The full line
 represents the mean light curve. The Blazhko effect is very pronounced
 around both minimum and maximum light. }
 \end{figure}

\begin{figure} 
\includegraphics[width=9cm]{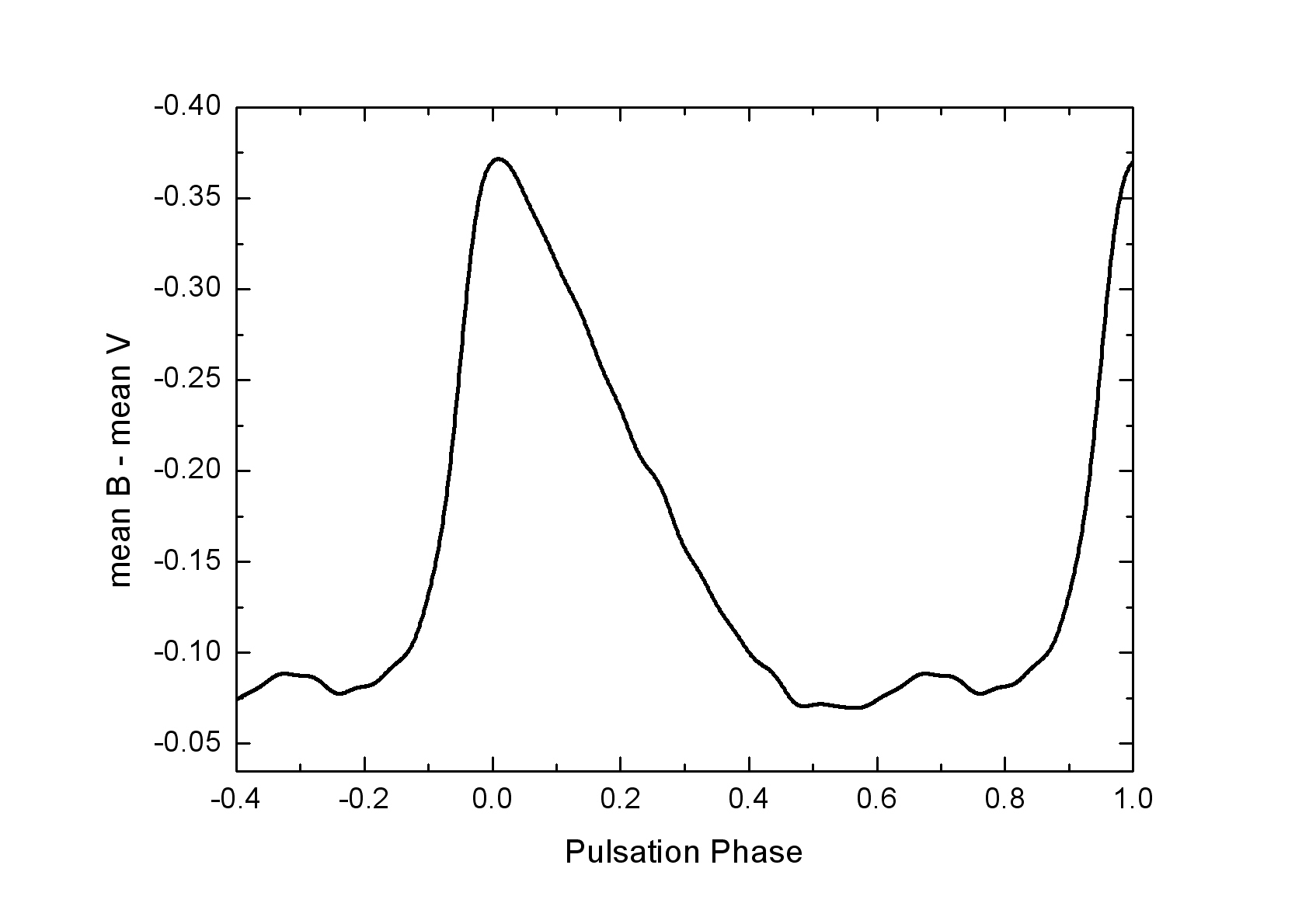}
 \caption{Mean $B-V$ color curve of SS For, resulting from subtracting the two mean light curves (Figure 3). }
 \end{figure}

\subsection{ASAS-3 data for SS For}

SS For (ASAS J020752-2651.9) was observed as part of the All Sky Automated Survey (ASAS-3; Pojmanski
2002) from November 2000 onwards. 
In most cases ASAS
photometry is accurate to about 0.03 mag (Pojmanski 2000).
We analysed SS For data from
ASAS-3 which were gathered between November 2000 and July 2008 (HJD
2451868.613--2454656.892, 484 points over 2788 days or almost 80 Blazhko
cycles). 
The folded $V$ ASAS-3 light curves are shown in Fig.\,4.

\begin{figure}
\includegraphics[width=9cm]{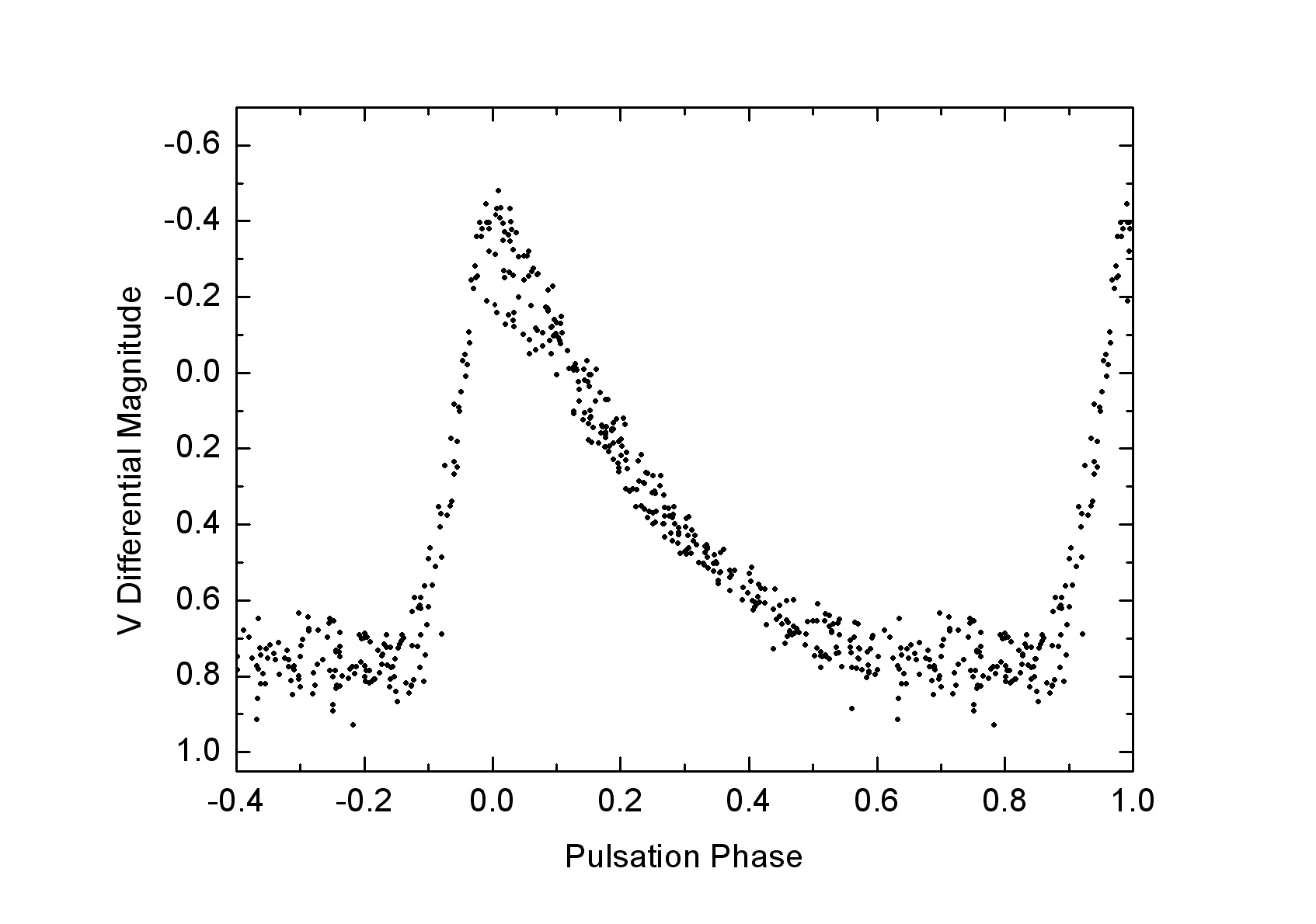}
 \caption{SS For data from ASAS-3, folded with the main pulsation period. }
\end{figure}

\section{Data analysis}

\begin{figure} 
\includegraphics[width=9cm]{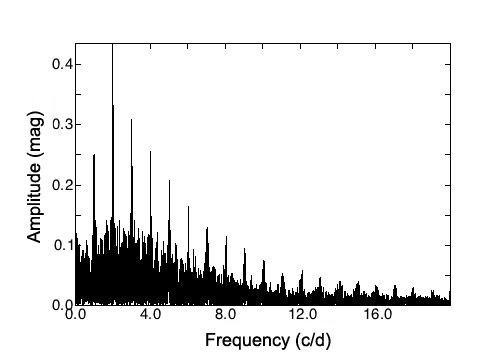}
 \caption{Fourier spectrum for the data set gathered at SAAO and SSO
 (2004-2005). The highest peak corresponds to the main frequency at 2.018433
 cd$^{\rm -1}$. The 1-day aliasing is clearly visible.}
\end{figure}

The frequency analyses were performed with Period04 (Lenz
\& Breger 2005), a package applying single-frequency power spectra and
simultaneous multi-frequency sine-wave fitting.  Fourier analyses were carried out on subsets of the photometric data in
both filters. Before merging them, we analysed each of the standardized data sets (SAAO, SSO, ASAS) separately to check the compatibility of the frequency solutions.

\subsection{Triplet fits: may be not sufficient}

The best fits to the data were determined by means of a
successive prewhitening strategy.  From this analysis the triplet structures
typical for RR Lyrae Blazhko stars clearly emerged.

For Blazhko stars very often the {\it a priori} choice is made to describe 
the light curve variations by means of {\it equidistant triplets} at the main
frequency and its harmonics according to the formula (see also Kov\'acs 1995):
\begin{equation}
\renewcommand{\arraystretch}{1.2}\begin{array}{l}
f(t) = A_0 + \sum_{k=1}^n [A_k \sin (2 \pi kf_0 t + \phi_k)\\
+ A_{k+} \sin (2 \pi (kf_0 + f_B)t + \phi_{k+})\\
+ A_{k-} \sin (2 \pi (kf_0 - f_B)t + \phi_{k-})]\\
+ B_0 \sin (2 \pi (f_B t + \phi_B)).
\end{array}
\end{equation}
The amplitudes and phases are calculated up to order $n$, for which the
amplitudes of the higher-order harmonics are still above the significance 
level.
In this formula, $f_0$ is the main pulsation frequency and $f_B$ is the Blazhko
frequency.  

In order not to presume equidistantly spaced triplets we also fitted the data with triplet structures without the condition of
equidistance, i.e. the side peak frequencies are "let free". The resulting departure from equidistance (see also Breger \& Kolenberg
2006) was smaller than the uncertainties on the frequencies, so we can resort to Equation (1) for the triplet fits.

We calculated optimum values for the 
frequencies, their amplitudes and phases by minimizing the residuals of the
fit given by Equation\,(1). 
Errors on these values
were determined through extensive Monte Carlo simulations, and are of the same
order as the error values calculated following Breger et al. (1999) and
Montgomery \& O'Donogue (1999).  
Handler et al. (2000) found that correlated
noise may lead to an underestimation of the errors.  Following their
conclusions we multiplied the errors resulting from our Monte Carlo simulations by a factor of 2, in order to
get a reliable and realistic error margin. The uncertainties obtained and given in Table\,2 were confirmed by our analyses of different subsets of the data.







\subsubsection{ASAS data}

With the ASAS-3 data set until July 2008 we find $f_0$=2.01843 $\pm$ 0.00001 cd$^{\rm -1}$ for the main frequency and $f_0+f_B$=2.04712 $\pm$ 0.00002 cd$^{\rm -1}$ for the first detected side peak frequency, resulting in a Blazhko frequency $f_B$=0.02868 $\pm$ 0.00002 cd$^{\rm -1}$.
Wils \& S\'odor (2005) published the results of a period
analysis of a set of southern RR Lyrae stars exhibiting the Blazhko effect and
published new elements for previously known as well as new and suspected 
Blazhko RRab stars of the ASAS database. They
determined the Blazhko periods of the variables with a fit containing the first four
trial modulation frequencies, i.e., a fit including the triplet components around the main frequency $f_0$ and its harmonic $2f_0$.
On the basis of the earlier ASAS-3 photometry for SS For, 
they list a Blazhko period of 34.8 d, which corresponds well with the
value we find.

We checked the ASAS data for period changes by looking at all normal maxima in the ASAS data.

\subsubsection{SAAO and SSO data}

\vspace{5mm}
\noindent
{\bf The $V$ data}
\vspace{5mm}

The highest recorded peak-to-peak amplitude in our $V$ data is 1.34 mag.
The harmonics of the main radial frequency are significant up to $16f_0$ and both triplet components are detected above the $3.5\sigma$ noise
level for combination frequencies up to the 11th order.  

When including the ASAS-3 $V$ data in the analysis, we obtain $f_0 = 2.01843 \pm 0.00001 $ 
cd$^{\rm -1}$ for the radial pulsation frequency and 
$f_N = 2.04708
\pm 0.00009 $ 
cd$^{\rm -1}$ for the right side peak frequency.  
Given the longer time span of the data set, we use the frequencies found from
the combined ASAS-3+SAAO+SSO data as {\it start values} for optimization to fit the new data gathered at SAAO and SSO. 
Note that we do not use the ASAS data to determine the final fit given in Table\,2.


\vspace{5mm}
\noindent
{\bf The $B$ data}
\vspace{5mm}

RR Lyrae stars have higher amplitudes in the blue region of the spectrum.
The highest recorded peak-to-peak amplitude from our $B$ data set is 1.65 mag.
For the main frequency and its harmonics, the amplitudes in $B$ are on average
a factor of about 1.23 larger than those in $V$ for $f_0$ and its harmonics up to order 10. 

Since they were based upon a larger and more extended set of
data, we used the optimized frequencies of the fit to the $V$ data to fit our
data obtained in the $B$ filter.  
Given the higher amplitudes in the $B$ filter, the harmonics of $f_0$ are
significant up to a higher order, $17f_0$.  Also both triplet components have amplitudes
above the $3.5\sigma$ noise level (the adopted threshold for combination frequencies) up to the 10th order. 

\vspace{5mm}
\noindent
{\bf The final fit}
\vspace{5mm}

Table\,2 list the results of a multifrequency fit to the 
combined set of SAAO+SSO data, according to Equation (1) and based on the magnitude differences
relative to HD13181, the comparison star with coordinates closer to SS
For and a smaller colour difference to the target. The results based on the
magnitude differences relative to HD13334 differ from the ones published in
Table\,2 within the given uncertainties. Note that for independent frequencies (in this case only $f_0$ and $f_B$) to be significant we required a signal-to-noise level exceeding 4. For combination frequencies we required the signal-to-noise level to be higher than 3.5.

\begin{table*}
\caption{Amplitudes and phases (fraction of $2\pi$) of the pulsation and modulation 
frequency components of SS For (triplets) for the best fit including triplet components.  The values displayed in {\it italics}
correspond to combination frequencies not exceeding a signal-to-noise level 
of 3.5.  The residuals of the triplet fit to the $V$ data are 0.017 mag, to the $B$ data
0.019 mag. At the bottom of the table we list the most prominent quintuplet components that are found in the data set. 
After inclusion of the 3 prominent quintuplet components the residuals are  0.015 mag both in $B$ and $V$.
The uncertainty on $f_0$ is  $10^{-6}$ cd$^{\rm -1}$ and the
uncertainty on $f_0+f_B$ is $4 \times 10^{-5}$ cd$^{\rm -1}$.  }
\centering                          

\begin{tabular}{llllllll}        
\hline               
\hline
\multicolumn{2}{c}{$f$ [cd$^{\rm -1}$]} &
\multicolumn{1}{l}{$A_V$ [mag]} &
\multicolumn{1}{l}{$\phi_V$ [$\frac{\rm rad}{2\pi}$]} &
\multicolumn{1}{l}{$\sigma(\phi_V)$} & 
\multicolumn{1}{l}{$A_B$ [mag]} &
\multicolumn{1}{l}{$\phi_B$ [$\frac{\rm rad}{2\pi}$]} &
\multicolumn{1}{l}{$\sigma(\phi_B)$ } \\
&& $\pm 0.003$ & & & $\pm 0.003$ & & \\
\hline                     
$f_0$   &      2.018433 &     0.424  &   0.995 & 0.001&0.550 &0.001 & 0.001\\
$2f_0$  &      4.036866 &    0.200  &   0.355 & 0.002& 0.248 &0.352 &0.002\\    
$3f_0$  &      6.055300 &    0.129  &   0.761 &0.003 & 0.159 & 0.758 & 0.003\\
$4f_0$  &      8.073733  &    0.080 &   0.174 & 0.006 &0.097 & 0.176 & 0.004\\
$5f_0$  &       10.092166  &   0.052 &     0.563 & 0.008 &0.063 &0.558 & 0.007\\  
$6f_0$  &       12.110600  &   0.039 &    0.970 & 0.011 & 0.047 &0.969 & 0.008\\
$7f_0$  &       14.129033  &  0.027 &   0.381 & 0.015 & 0.034 &0.376 & 0.012\\
$8f_0$  &       16.147466  &   0.019 &   0.779 & 0.022 & 0.024 &0.783 & 0.018\\
$9f_0$  &       18.165899  &  0.013 &   0.185 &  0.034 & 0.015 &0.163 & 0.028\\
$10f_0$ &       20.184333  &  0.011 &    0.557 & 0.038 &0.013 & 0.565 & 0.033\\
$11f_0$ &     22.202766  & 0.008 &  0.988 & 0.054 &
	  0.009 & 0.990 & 0.047\\
$12f_0$ &     24.221199  & 0.006 &  0.384 & 0.066 &0.007 & 0.387 &0.053\\
$13f_0$ &     26.239632  & 0.005 & 0.772 & 0.088 &0.005 & 0.794 &0.081\\
$14f_0$ &     28.258066 & 0.003 &   0.099 & 0.120 &0.004 & 0.103&0.098\\
$15f_0$ &     30.276499  & 0.003 &    0.634 & 0.144 &0.003 & 0.622&0.148\\
$16f_0$ &  32.295048 & 0.002  & 0.009 &  0.140 &  0.002 &    0.982& 0.151\\       
$17f_0$ &  32.295048 & {\it 0.002}  & {\it 0.254} &  0.168 &  0.002 &    0.292 & 0.155\\  

$f_0+f_B$  &       2.04705 &   0.050   &  0.545       &0.007&0.061 &0.541 &0.006\\    
$f_0-f_B$  &       1.98982 &    0.024  &  0.399       &0.014&0.029 &0.403 &0.014\\
$2f_0+f_B$  &      4.06548 &    0.006 &   0.820 &0.060 & 0.005 &
  0.874 & 0.090\\ 
$2f_0-f_B$  &      4.00825 &      0.014  &  0.642 & 0.030
&0.019 &0.621&0.022\\
$3f_0+f_B$  &      6.08392 &  0.009  &  0.226  & 0.038 & 0.011& 
  0.215 & 0.033\\
$3f_0-f_B$  &      6.02668 &     0.017  &  0.036  & 0.020 & 0.020 & 0.043 &0.020\\
$4f_0+f_B$  &     8.10235 &    0.020   & 0.583   &0.018 & 0.025 &0.583&0.016\\
$4f_0-f_B$  &     8.04512 &     0.012 &   0.462  &  0.031
&0.015&0.458 &0.026\\
$5f_0+f_B$  &      10.12078 &     0.015  &  0.050  &0.025   &0.018 &0.039 &0.021\\
$5f_0-f_B$  &      10.06355 &  0.013 & 0.945 &  0.026 &0.016&0.950&0.024\\
$6f_0+f_B$  &       12.13922 &     0.011  &  0.477 &0.033 & 0.013 &0.464&0.028\\
$6f_0-f_B$  &        12.08198 &  0.010 & 0.345 &  0.039  &0.013&0.339&0.027\\
$7f_0+f_B$  &       14.15765 &     0.009  &  0.874 &0.042 &  0.011 &0.859&0.038\\
$7f_0-f_B$  &       14.10041 &  0.008 &    0.793 &  0.042 &0.011&0.805&0.040\\
$8f_0+f_B$  &       16.17608 &     0.006  & 0.247 & 0.061&0.008&0.256&0.048\\
$8f_0-f_B$  &        16.11884 &    0.006 & 0.273 & 0.060 &0.006&0.282&0.070\\
$9f_0+f_B$  &       18.19451 &     0.005  & 0.694 & 0.074&0.005&0.681&0.081\\
$9f_0-f_B$  &       18.13728 &    0.005 &  0.644 & 0.082 &0.007&0.654&0.052\\
$10f_0+f_B$  &      20.21295 &     0.003 &  0.085 & 0.118 &0.005&0.048&0.073\\
$10f_0-f_B$  &       20.15571 &    0.004 &   0.063 & 0.088 &0.005&0.106&0.081\\
$11f_0+f_B$  &       22.23138 &   0.003  & 0.426 & 0.105&0.004&0.452&0.101\\
$11f_0-f_B$  &        22.17415 &   0.003 &   0.554 & 0.116 & {\it 0.003} & {\it 0.496} & {\it 0.140}\\
$12f_0+f_B$  &       24.24981 &     0.003  &  0.895 &0.124& {\it 0.003} & {\it 0.827} & {\it 0.141}\\
$12f_0-f_B$  &        24.19258 &   {\it 0.001} &  {\it 0.939} &{\it 0.330
}&{\it 0.003} & {\it 0.008} & {\it 0.146} \\
$13f_0+f_B$  &       26.26825 &    0.003  & 0.299 & 0.141&0.003 & 0.319 &0.122\\
$13f_0-f_B$  &       26.21101 &   {\it 0.001} &  {\it 0.312} &{\it 0.366
}&{\it 0.002} & {\it 0.279} & {\it 0.320} \\
$14f_0+f_B$  &       28.28668 &    {\it 0.002}  & {\it 0.763} &{\it 0.325}&0.001&0.769 &0.303\\
$14f_0-f_B$  &      28.22945 &   0.001 &  0.728 & 0.152
& 0.003 & 0.718 & 0.129 \\
$15f_0+f_B$  &      30.30511   &     0.002  &  0.042 &
  0.156  & 0.003 & 0.045 & 0.128\\
$15f_0-f_B$  &      30.24788 &   {\it 0.003} &  {\it 0.126} &{\it 0.408
}&{\it 0.001} & {\it 0.099} & {\it 0.276} \\
$f_B$  &   0.028617 & 0.007 & 0.907 & 0.048 & 0.006
	& 0.887 & 0.051\\
\hline
\multicolumn{8}{l}{Most prominent quintuplet components found in the data:}\\
\multicolumn{8}{l}{}\\

$3f_0+2f_B$ & 6.1125 &   0.008 & 0.54 &   0.04 & 0.014
	&  0.85 & 0.03\\
$6f_0+2f_B$ & 12.1678 &  0.008 & 0.88 & 0.04 & 0.011 
	& 0.97 & 0.04\\
$5f_0-2f_B$ & 10.0349 &  0.008 & 0.26 & 0.03 & 0.009
	& 0.33 & 0.03\\
	... & ... &  ... & ... & ... & ...
	& ... & ...\\
			
\hline
\end{tabular}
\end{table*}

\vspace{5mm}
\noindent
{\bf Ephemerides}
\vspace{5mm}

From the new SAAO and SSO data we derived the following ephemerides of maximum
light and maximum pulsation amplitude:  

\begin{equation}
{\rm HJD} (T_{\rm max}) = 2453296.440 + 0.495433  \times E_{\rm pulsation}
\end{equation}
\begin{equation}
{\rm HJD} (T_{\rm Blmax}) = 2453298.914 + 34.94 (\pm 0.05) \times E_{\rm Blazhko}\\
\end{equation}

The time of maximum light $(T_{\rm max})$ was recorded in our data.  The time of the highest light curve amplitude 
$(T_{\rm Blmax})$ was determined on the basis of our best fit to the data.  Our highest recorded maximum was at HJD=2453296.440 but according to our best fit it preceded the absolute maximum, which happened 5 pulsation cycles later at HJD=2453298.914.

\subsection{Residual power and quintuplet components}

\begin{figure*}
\includegraphics[width=160mm, angle=0]{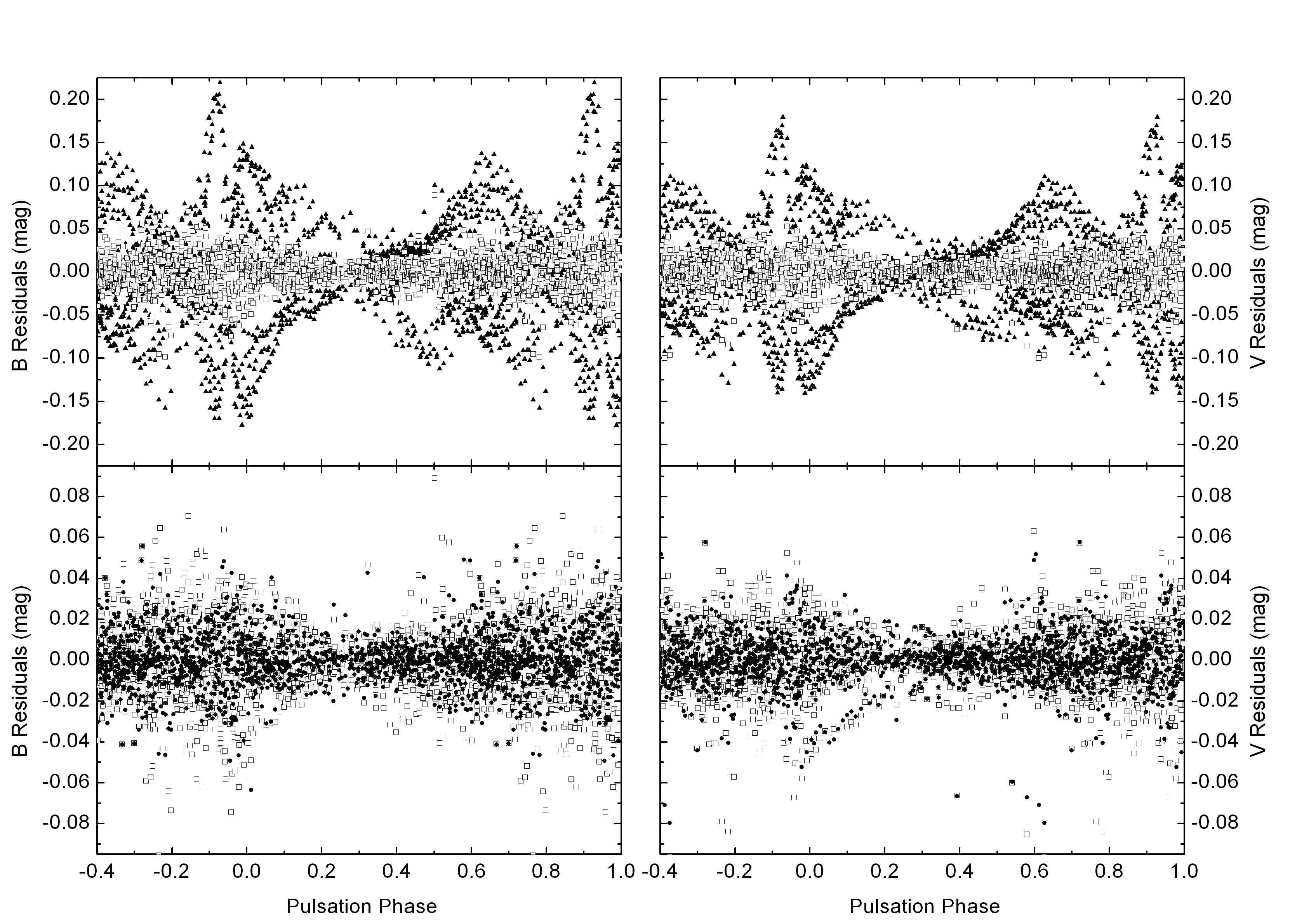}
 \caption{Residuals of the $B$ (left panel) and $V$ (right panel) data of SS For after subtraction of the mean
 light curve (top panels, open circles) and after subtraction of the best fit including triplets (top and bottom panels, filled grey circles).  When including the quintuplet components in the fit (bottom panels with a larger magnitude scale, filled black circles), the residuals reduce even more.}
\end{figure*}

\begin{figure}
\includegraphics[width=95mm]{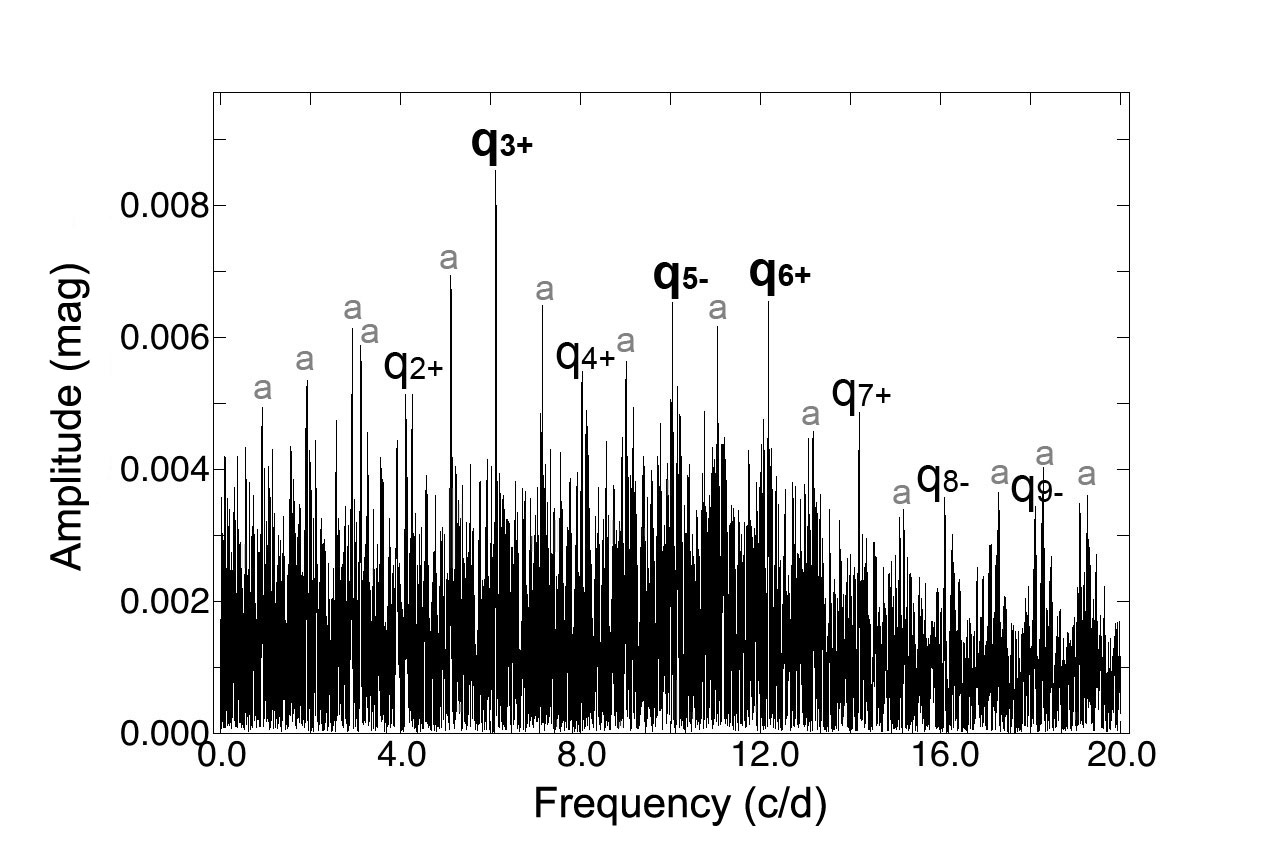}
 \caption{Fourier spectrum showing the quintuplet components found in the data for SS For after subtraction of the main frequency. its harmonics and the triplet components. The notation "$q_{k+}$" and "$q_{k-}$" stands for $kf_0+2f_B$ and $kf_0-2f_B$ respectively; "a" denotes alias peaks. }
\end{figure}

After subtraction of the best triplet fit to the data, there is still residual scatter
of the light curve (Figure\,6, grey filled circles), concentrated at the phases of minumum to maximum light (pulsation phase $\phi = 0.8-1.1$) and in the bump phase (pulsation phase $\phi = 0.5-0.8$). This indicates that there are most likely other periodicities present in the data set.  Therefore, we searched for additional frequencies in the residuals after subtraction of the fit including the main frequency, its harmonics and all detectable triplet components (see Table\,2).  

Applying this procedure to the $B$ data in which the modulation components have the higher amplitudes, we first detected a peak at the frequency 6.1125 cd$^{\rm -1}$ with a significant amplitude. This peak is located at the expected position of the higher quintuplet component at 3$f_0$, $3f_0+2f_B$. We also detected significant power at the following frequencies: 12.1678 cd$^{\rm -1}$ ($6f_0+2f_B$) and 10.0349 cd$^{\rm -1}$ ($5f_0-2f_B$).

separations within the uncertainty on the frequency. As it is quite remarkable to find additional frequencies exactly at the expected positions of quintuplet components, we have added the found frequencies, their amplitudes and phases below Table\,2.  Figure\,7 shows the Fourier spectrum in the $B$ data after subtraction of the triplet solution.  Quintuplet components and aliases are indicated. We clearly see even more quintuplet components in the Fourier spectrum than were obtained through subsequent prewhitening.
The quintuplet components emerge most clearly from the $B$ data set, but we find the same frequencies with significant amplitudes in the $V$ data.  

\begin{figure}
\includegraphics[width=9cm]{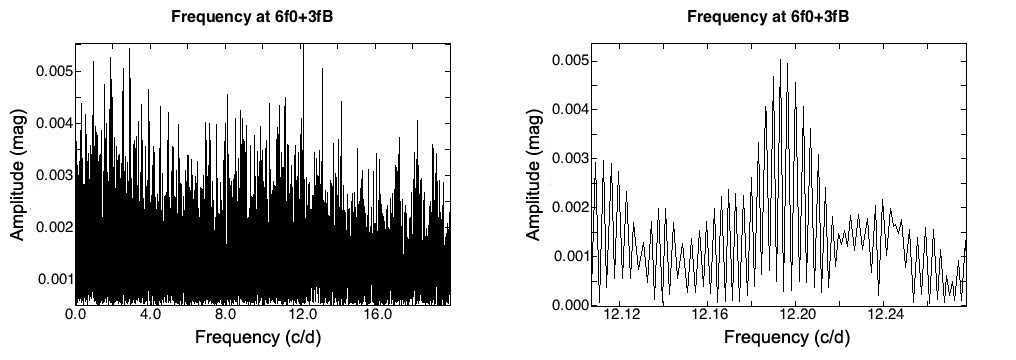}
 \caption{Fourier spectrum for an additional frequency component found in the $B$ data after subtraction of the triplet fit plus the found quintuplet components.  Left panel: Spectrum between 0-20  cd$^{\rm -1}$, right panel: zoom. The peak is at the expected position of a septuplet component  ($6f_0+3f_B$).}
\end{figure}

After including the detected quintuplet components in the fit, we find residual power at several frequencies around the significance level. None of them is clearly a quintuplet component and at this level spurious peaks occur.  More quintuplet components may be present in the data, but were not unambiguously detected by us. Given the imperfect coverage of our data (gaps) we considered it inappropriate to fit more quintuplet components to the data.  We note, however, that in the $B$ data we find evidence for a significant peak at 12.1964 cd$^{\rm -1}$, corresponding to ($6f_0+3f_B$, see Figure\,8).  More evidence is needed to confirm this frequency as one really present in the data and not an artefact of the sampling.  Jurcsik et al. (2008) identified frequencies at $kf_0 \pm 4f_B$ in data of MW Lyr.  They also found septuplet frequencies (i.e., frequencies at $kf_0 \pm 3f_B$) and  $kf_0 \pm 5f_B$ components in the residuals of their data set, though much less significantly.  Multiplet components $kf_0 \pm jf_B$ with $j \ge 3$ (i.e., septuplet, nonuplet, etc. components) have not been considered in any of the current models for explaining the Blazhko effect. 

Table\,2 gives the triplet components, as these produce a stable fit and hence are suited for calculations of derived quantities (with appropriate error bars). We also applied a fit including equidistant quintuplet components to the data set, but found this fit to be unstable. For the main frequency components (including harmonics) there is no significant difference between the triplet and quintuplet fit.  In the fit including all quintuplet components, however, some side peak amplitudes and phases may drift because of the high number of free parameters and close frequencies. 
  
As shown in Figure\,6, the scatter in the residuals decreases when the quintuplet frequencies are taken into account. 
The new fit including the detected quintuplet components makes it possible to more accurately describe the variations at the bump phase and around minimum and maximum light.


\subsection{Properties of the modulation components}

The properties of the modulation components $kf_0+f_B$ and $kf_0-f_B$ in the
frequency spectra of a Blazhko star constitute an important test for the
models proposed to explain the modulation.  
However, it is important to obtain reliable uncertainties for the amplitudes and phases of the modulation components.  As was shown by Jurcsik et al. (2005a), the amplitudes of the triplet (or generally: multiplet) components depend strongly on the coverage.
 
\subsubsection{Amplitude ratios, phase differences and asymmetry parameters}

From their extensive study comprising 731 Blazhko variables in the LMC, Alcock et
al. (2003) found that the relative amplitudes of the first order modulation
components are usually in the range $0.1<A_{\pm}/A_1<0.3$, but there are some
extreme values.  For our $V$ data, we find
$A_+/A_1 = 0.118 \pm 0.007$ and
$A_-/A_1 = 0.057 \pm 0.007$.
For the $B$ data, 
$A_+/A_1 = 0.111 \pm 0.005$ and
$A_-/A_1 = 0.053 \pm 0.005$.
 
Table\,3 lists the amplitude ratios, phase differences and their errors for
the first six modulation component pairs, defined as
\begin{equation}
R_k = A_{kf_0+f_B}/A_{kf_0-f_B}
\end{equation}
and
\begin{equation}
\Delta \phi_k = \phi_{kf_0+f_B}-\phi_{kf_0-f_B}.
\end{equation}
We also list the so-called
asymmetry parameter 
\begin{equation}
Q=\frac{A_{+}-A_{-}}{A_{+}+A_{-}},
\end{equation}
introduced by Alcock et al. (2003) to quantify the degree of asymmetry in the peaks.
The distribution of the $Q$ parameter for the
Blazhko stars from the MACHO data base (Alcock et al. 2003) peaks at +0.3. 

According to
Kov\'acs (1995, and references therein), the parameter combinations $R_k$ and $\Delta
\phi_k$ are expected to be constant under the assumption of a simple oblique
rotator-pulsator model. However, such a simple model seems inadequate to
explain the variety of observations related to the Blazhko effect, and hence
we give Table\,3 rather as light curve diagnostics (see also Smith
et al. 1999 and Jurcsik et al. 2005a).  

Due to the rather large error bars on
the parameter combinations, the phase
differences $\Delta \phi_k$, especially for the $V$ data, do not deviate significantly from a constant value. On the other hand, the amplitude ratios $R_k$ show
large changes, both in $V$ and in $B$, and with the obtained error bars they can definitely not be considered as constant.

As can be seen in Table\,3 the $Q$ value can change its sign in different orders.
The positive $Q$ values at multiplet order $k = 1$, 4 and 5 point to an asymmetry with higher amplitudes at the higher
frequency lobes in the triplets, as is mostly the case in Blazhko stars.  At multiplet orders $k$ = 2 and 3 the left side peaks are higher, yielding negative $Q$ values. 

\begin{table*}
\caption{Side lobe amplitude ratios $R_k$, phase differences $\Delta \phi_k$
(fraction of $2\pi$) and asymmetry
  parameters $Q$ as defined in the
  text, and their respective
  errors for the $V$ and $B$ data.  
$k$ denotes the multiplet order. }             
\label{table:1}      
\centering                          

\begin{tabular}{c c c c c c c c c c c c c}        
\hline
\hline               
$k$ & $R_k(V)$ & $\sigma_{R_k}(V)$ & $\Delta \phi_k(V)$ & $\sigma_{\Delta \phi_k}(V)$ &
$Q(V)$ & $\sigma_Q(V)$ & $R_k(B)$ & $\sigma_{R_k}(B)$ & $\Delta \phi_k(B)$ & $\sigma_{\Delta \phi_k}(B)$ &
$Q(B)$ & $\sigma_Q(B)$ \\
\hline      
1	&	2.05	&	0.28	&	0.15	&	0.02	&	0.34	&	0.07	&	2.09	&	0.24	&	0.14	&	0.02	&	0.35	&	0.06	\\
2	&	0.43	&	0.24	&	0.18	&	0.07	&	-0.40	&	0.31	&	0.26	&	0.16	&	0.25	&	0.09	&	-0.58	&	0.50	\\
3	&	0.53	&	0.19	&	0.19	&	0.04	&	-0.30	&	0.16	&	0.54	&	0.17	&	0.17	&	0.04	&	-0.30	&	0.14	\\
4	&	1.69	&	0.51	&	0.12	&	0.04	&	0.26	&	0.11	&	1.68	&	0.39	&	0.12	&	0.03	&	0.25	&	0.08	\\
5	&	1.18	&	0.36	&	0.11	&	0.04	&	0.08	&	0.04	&	1.08	&	0.27	&	0.09	&	0.03	&	0.04	&	0.01	\\
6	&	1.13	&	0.46	&	0.13	&	0.05	&	0.06	&	0.03	&	1.00	&	0.32	&	0.12	&	0.04	&	0.00	&	0.00	\\
\hline
\end{tabular}
\end{table*}

\subsubsection{Modulation in $B$ versus $V$}

The Blazhko modulation in the $B$ filter is larger than in the $V$ filter. 
Jurcsik et al. (2005b) found, from a combination of different data sets, 
that the value of $A_{\rm mod}(B)/A_{\rm mod}(V)$ lies within the range of 1.23
-- 1.39 with a mean value of 1.30.  Their value was determined from the sum of
the Fourier amplitudes of the first four modulation components, 
$A_{f_0+f_B}, A_{f_0-f_B}, A_{2f_0+f_B}, A_{2f_0-f_B}$.
For our data we obtain a value  $A_{\rm mod}(B)/A_{\rm mod}(V)=1.22 \pm 0.10$,
taking into account the first four modulation components. 
At
 $2f_0+f_B$ the side peak amplitude determined from our data set is rather low.
When the higher order side peaks are taken into account the value of $A_{\rm
  mod}(B)/A_{\rm mod}(V)$ does not increase significantly (see Table\,4).
This implies that the modulation is indeed stronger in $B$ than in $V$, and by the same factor as holds for the main frequency and its harmonics (Section 3.2.1).

\begin{table} 
\caption{The ratio of the strength of the modulation components in the Johnson $B$
and $V$ filter for increasing order $k$.  $k=1$ takes into account the side
peak amplitudes around $f_0$,...,$kf_0$. }             
\centering                          

\begin{tabular}{c c c c}        
\hline
\hline  
Order $k$ & $A_{\rm mod}(B)$ & $ A_{\rm mod}(V)$ & $A_{\rm mod}(B)/A_{\rm mod}(V)$\\
\hline																	
1	&	0.091	$\pm$	0.004	&	0.075	$\pm$	0.004	&	1.21	$\pm$	0.09	\\																
2	&	0.115	$\pm$	0.006	&	0.094	$\pm$	0.006	&	1.22	$\pm$	0.10	\\																		
3	&	0.146	$\pm$	0.007	&	0.121	$\pm$	0.007	&	1.20	$\pm$	0.09	\\																		
4	&	0.185	$\pm$	0.008	&	0.152	$\pm$	0.008	&	1.21	$\pm$	0.09	\\																		
5	&	0.219	$\pm$	0.009	&	0.180	$\pm$	0.009	&	1.22	$\pm$	0.08	\\																		
6	&	0.246	$\pm$	0.010	&	0.201	$\pm$	0.010	&	1.22	$\pm$	0.08	\\																		
7	&	0.268	$\pm$	0.011	&	0.219	$\pm$	0.011	&	1.22	$\pm$	0.08	\\																		
8	&	0.282	$\pm$	0.012	&	0.231	$\pm$	0.012	&	1.22	$\pm$	0.08	\\																		
9	&	0.294	$\pm$	0.013	&	0.241	$\pm$	0.013	&	1.22	$\pm$	0.08	\\																		
10	&	0.304	$\pm$	0.013	&	0.248	$\pm$	0.013	&	1.23	$\pm$	0.09	\\
\hline																		
\end{tabular}
\end{table}

\subsubsection{Amplitude decrease for the side peaks}
Jurcsik et al. (2005a) reported that the decrease in amplitude of the
successive harmonics of $f_0$ is much steeper than that over the side peaks.
Whereas the latter decrease can be described as exponential, the side peak
amplitudes have a more linear decrease.  
As a consequence, the relative contribution of the modulation at higher orders is larger.
This is illustrated in Figures\,9
and 10, showing the amplitude ratios $A(f_k)/A(f_0)$,  $A(f_k+f_B)/A(f_0+f_B)$, and
 $A(f_k-f_B)/A(f_0-f_B)$ for the $V$ and the $B$ data respectively.
 The amplitudes of the left and right
 side peaks decrease less steeply. They also show more irregular behavior at lower multiplet orders ($k = 2, 3, 4$).

\begin{figure} 
\includegraphics[width=95mm, angle=0]{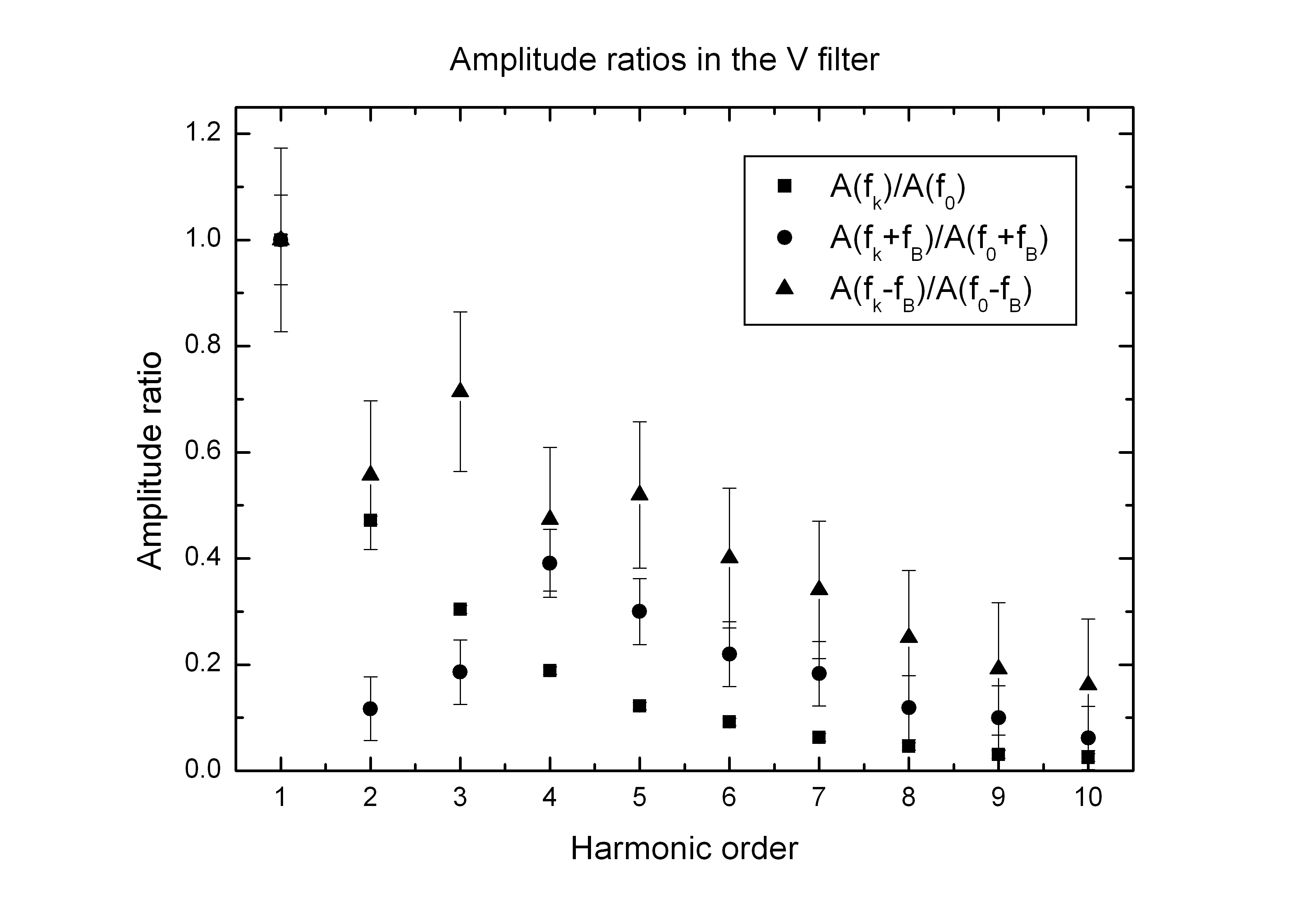}
 \caption{Amplitude ratios $A(f_k)/A(f_0)$,  $A(f_k+f_B)/A(f_0+f_B)$, and
 $A(f_k-f_B)/A(f_0-f_B)$ of the detected frequencies for the $V$ data. The amplitudes of the main frequency and its harmonics decrease
 exponentially with increasing order.  The amplitudes of the left and right
 side peaks, on the other hand, decrease less steeply and show more irregular behavior at lower multiplet orders. }
\end{figure}
\begin{figure} 
\includegraphics[width=95mm, angle=0]{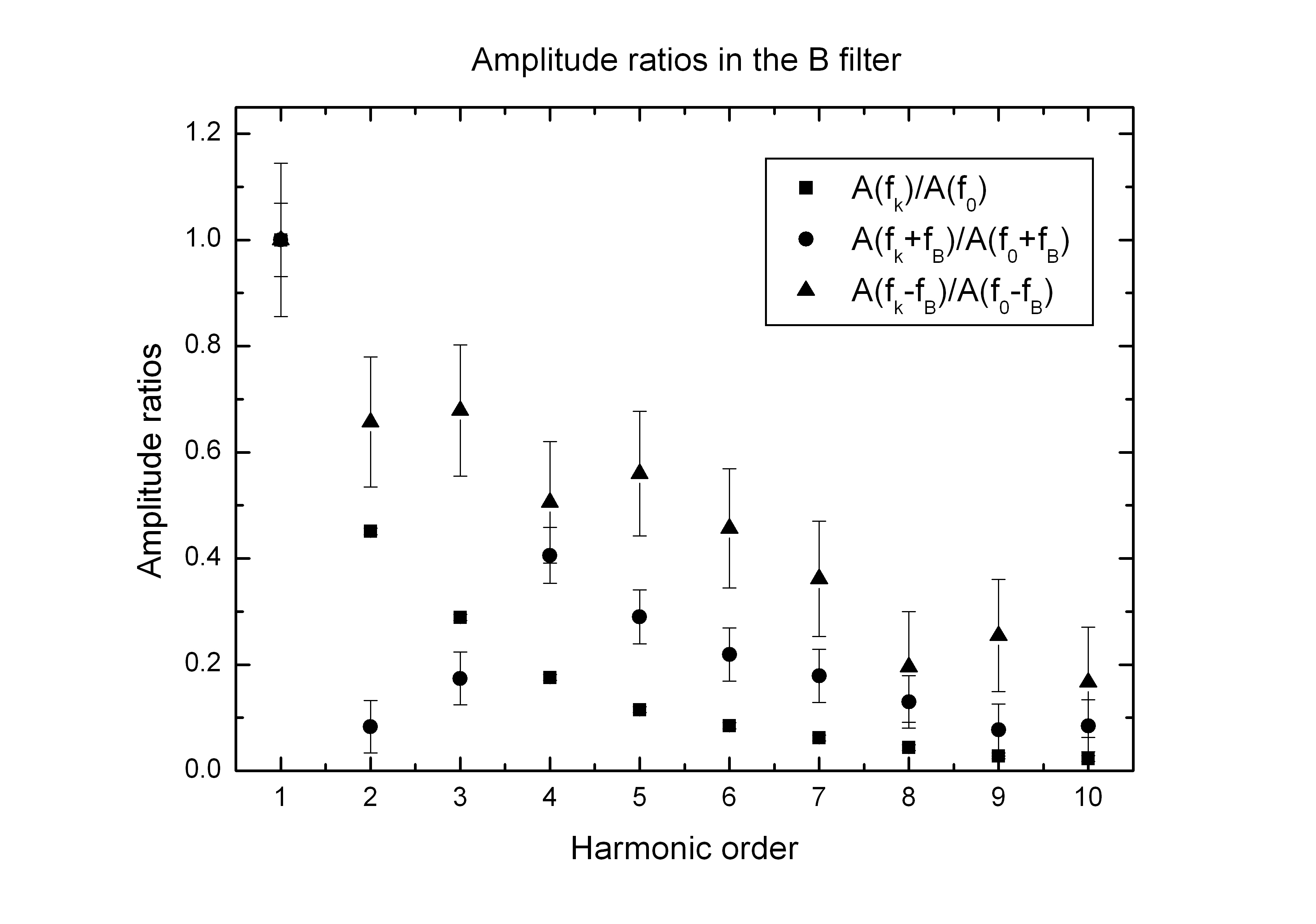}
 \caption{Amplitude ratios $A(f_k)/A(f_0)$,  $A(f_k+f_B)/A(f_0+f_B)$, and
 $A(f_k-f_B)/A(f_0-f_B)$ of the detected frequencies for the $B$ data. See also
 caption of Figure\,9.}
\end{figure}

\subsubsection{Light curve modulation}
Figure\,6 shows the residual light curve of SS For folded with the main
pulsation period after subtraction of the mean light curve (open circles).  It
shows that the Blazhko modulation in SS For is not only concentrated in the
minimum to maximum phase interval, as is the case for other Blazhko stars
(see, e.g., Jurcsik 2005a).  The strong variations in the bump phase before
minimum light ($\phi=0.5-0.8$, see also Guggenberger \& Kolenberg 2006) are
reflected in the scatter.  
As is known for
Blazhko variables, the scatter is generally largest around the phase of maximum light.
For SS For, there is a comparable scatter around minimum light.
A large modulation around minimum light was also detected in the Blazhko star MW Lyr (Figure\,3 in Jurcsik et al. 2008).

\subsection{Variations over the Blazhko cycle}

In order to assess the light curve variations over the Blazhko cycle, our data
were divided into 10 subsets.  Each subset corresponds to a 0.1 phase interval
over the Blazhko cycle.  For the ephemeris of Blazhko phase $\Psi=0$, defined
as the phase of the brightest maximum, we adopted the time of the highest 
detected maximum obtained from the best fit to our observations, i.e., $T_0= 245 3298.914$.

\begin{table*} 
\caption{A log of the subsets of the observations used to construct the light
  curves at different phase intervals in the Blazhko cycle. For the $V$ data set we used SAAO, SSO and ASAS data, for the $B$ filter only SAAO and SSO.}             
\centering                          
\begin{tabular}{c c c c c c c}       
\hline
\hline              
$\Delta\Psi$ & HJD (-2450000) $(V)$ & $N(V)$ & HJD (-2450000) $(B)$ & $N(B)$\\
\hline                       
0.0-0.1	&	1868.61332	--	4629.92251	&	201 & 3296.440--3612.133 & 132\\
0.1-0.2	&	1870.60972	-	4527.50794	&	248 & 3300.380--3581.324 & 184\\
0.2-0.3	&	1873.60768	-	4633.91429	&	85 & 3303.413--3584.678 & 174\\
0.3-0.4	&	1878.59277	-	4534.49935	&	84  & 3587.590--3622.167 & 38\\
0.4-0.5	&	1880.57795	-	4501.57064	&	133  & 3590.204--3591.321 & 24\\
0.5-0.6	&	1884.56548	-	4645.90353	&	163 & 3592.519--3594.678 & 96\\
0.6-0.7	&	1888.56074	-	4649.89091	&	241 & 3319.402--3633.294 & 137\\
0.7-0.8	&	2067.93681	-	4652.92994	&	209 & 3288.278--3602.671 & 204\\
0.8-0.9	&	1930.5492	-	4656.89163	&	110 & 3289.481--3571.323 & 158\\
0.9-1.0	&	1900.53501	-	4520.53379	&	206  & 3296.257--3574.686 & 66\\
\hline                                   
\end{tabular}
\end{table*}

\begin{figure*}
\includegraphics[width=12cm, angle=0]{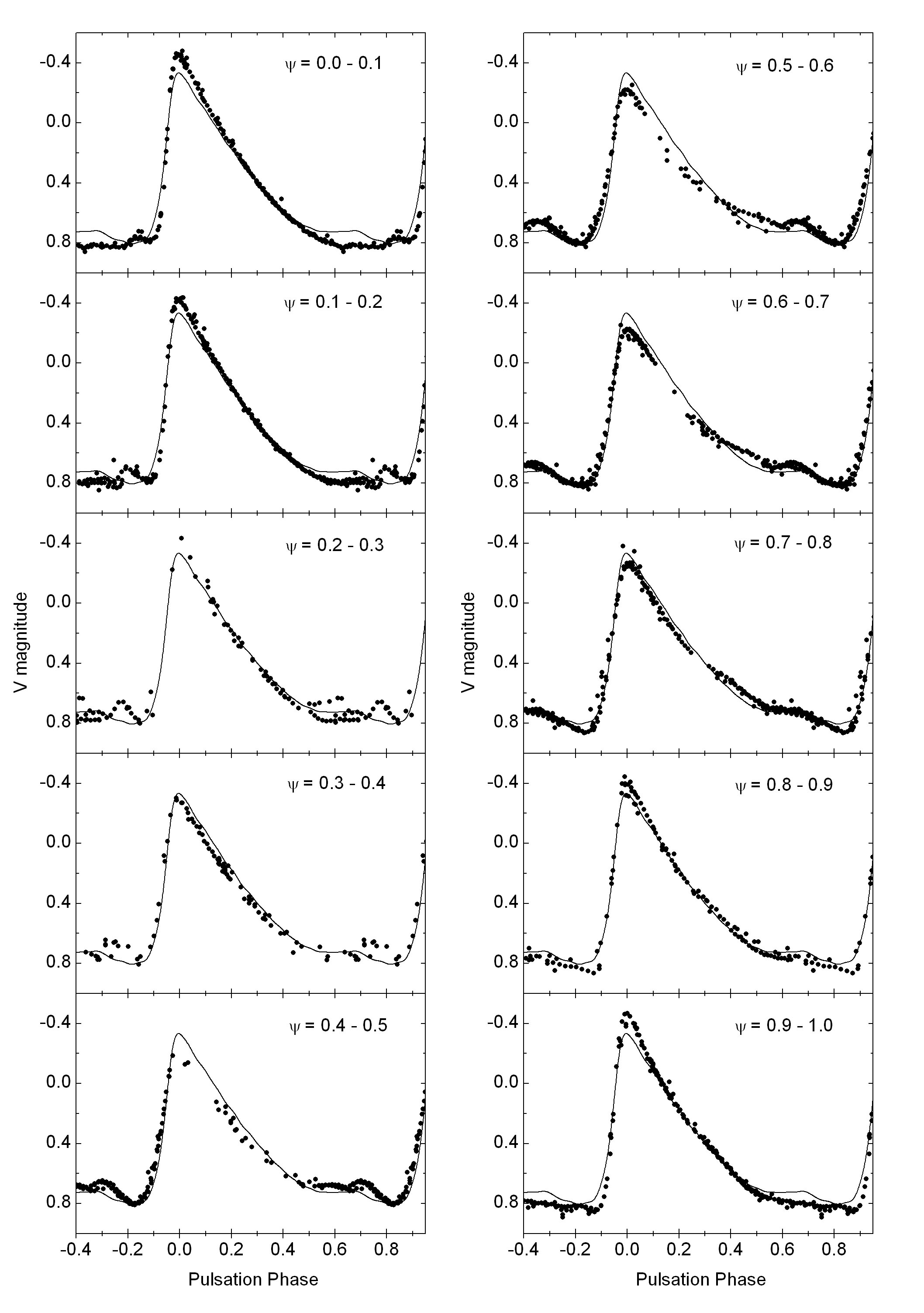}
 \caption{$V$ light data (crosses) in different 0.1 phase intervals of the
 Blazhko cycle (34.94 $\pm$ 0.05 d from this data set).  We used the SAAO, SSO and ASAS data for constructing the phase bins. The mean light curve over all Blazhko phase
 intervals is also shown as a solid line.  Note the strong variations at the
 bump phase before minimum light. }
\end{figure*}

Table\,5 gives the time base for each of these intervals, as well as the number
of data points.  
For the $B$ data the gaps in the light curves at different Blazhko phases do not allow us to
perform a fit for each of the phase intervals (Figure\,14, available online).  If coverage of minimum or
maximum light is lacking, it is impossible to reliably fit the data and determine Fourier parameters.
For the $V$ data, however, we added the ASAS-3 photometry, and fitted each of the
phase intervals with
a higher-order harmonic fit according to:
\begin{equation}
f(t) = A_0 + \sum_{k=1}^8 A_k \sin (2 \pi kf_k t + \phi_k),
\end{equation}
and
calculated Fourier parameters, namely the amplitude ratios
\begin{equation}
R_{k1} = A_{k}/A_{1}
\end{equation}
and the epoch-independent phase differences
\begin{equation}
\phi_{k1} = \phi_{k}-k\phi_{1}.
\end{equation}
These parameters offer a way to quantify the shape of the lightcurves.
Fig.\,12 shows the variations of the amplitudes $A_k$ and the phases
$\phi_k$ for $k=1,...,4$, and Figure\,13 the derived Fourier parameters.  
The Fourier parameters are also listed in Table\,6.
It is clear that the first order amplitude $A_1$ 
decreases towards minimum Blazhko amplitude more drastically than do the higher
order amplitudes $A_2$, $A_3$, $A_4$.
As a consequence, the $R_{k1}$ parameters reach their maximum amplitude around
minimum light.
Small variations in the $\phi_k$, and subsequently the $\phi_k1$ parameters, can be observed (Figure\,12). The variation of $\phi_1$ is the largest, as was also noted by Jurcsik et al. (2008).   

\begin{table*} 
\caption{Variations of the Fourier parameters $A_1$, $R_{k1}$ ,  $\phi_1$, and $\phi_{k1}$ over
 the Blazhko cycle for  $k=1,...,4$.}             
\centering                          
\begin{tabular}{c | c c c c c c c c}       
\hline
\hline      
$\Delta \Psi$	&	$A_1$ & $R_{21}$	&	$R_{31}$	&		$R_{41}$	&  $\phi_1$ &	$\phi_{21}$		&	$\phi_{31}$		&	$\phi_{41}$\\
\hline																									
0.0-0.1	&	0.502 $\pm$ 0.006 & 0.43	 $\pm$ 0.02	&	0.27		 $\pm$	0.01	&	0.19		 $\pm$	0.01	& -0.05	$\pm$ 0.01 &	2.28		 $\pm$	0.05	&	4.78		 $\pm$	0.07 &	1.07	$\pm$	0.09	\\
0.1-0.2	&	0.480 $\pm$ 0.006 & 0.42	 $\pm$	0.01	&	0.28		 $\pm$	0.01	&	0.20		 $\pm$	0.01	& -0.01 $\pm$ 0.01 &	2.18		 $\pm$	0.04	&	4.64		 $\pm$	0.06	&	0.91		 $\pm$ 	0.08	\\
0.2-0.3	&	0.450 $\pm$ 0.017 & 0.43		 $\pm$	0.03	&	0.29		 $\pm$	0.03	&	0.20		 $\pm$	0.03	& 0.06 $\pm$	0.03 &	2.10	 $\pm$	0.10	&	4.58		 $\pm$	0.13	&	0.75		 $\pm$	0.20	\\
0.3-0.4	&	0.411 $\pm$ 0.009 & 0.41		 $\pm$	0.03	&	0.29		 $\pm$	0.03	&	0.22		 $\pm$	0.03	&	0.07 $\pm$ 0.03 & 2.11		 $\pm$	0.08	&	4.60	 $\pm$	0.12	&	0.92		 $\pm$ 0.16	\\
0.4-0.5	&	0.340 $\pm$ 0.008 & 0.46	 $\pm$	0.04	&	0.34	 $\pm$	0.05	&	0.25		 $\pm$	0.03	&	0.04	$\pm$ 0.04 & 2.23		 $\pm$	0.11	&	5.06		 $\pm$	0.14	&	1.42		 $\pm$	0.22	\\
0.5-0.6 &	0.352 $\pm$ 0.007 & 0.55		 $\pm$	0.02	&	0.34		 $\pm$ 0.02	&	0.17	 $\pm$	0.02 & 0.04 $\pm$ 0.02	&	2.26		 $\pm$	0.06	&	4.86		 $\pm$	0.08	&	1.26		 $\pm$	0.14	\\
0.6-0.7	&	0.374 $\pm$ 0.005 & 0.54		 $\pm$	0.02	&	0.30		 $\pm$	0.02	&	0.15		 $\pm$	0.02	&	-0.04 $\pm$	0.02 & 2.36		 $\pm$	0.06	&	4.93		 $\pm$	0.08	&	1.58		 $\pm$	0.14	\\
0.7-0.8	&	0.402 $\pm$ 0.009 & 0.50	 $\pm$	0.02	&	0.33		 $\pm$	0.02	&	0.16		 $\pm$	0.02	&	-0.09 $\pm$ 0.02 & 2.49		 $\pm$	0.06	&	5.13		 $\pm$	0.08	&	1.62		 $\pm$ 0.15	\\
0.8-0.9	&	0.458 $\pm$ 0.010 & 0.46		 $\pm$	0.02	&	0.31		 $\pm$	0.02	&	0.18		 $\pm$	0.03	&	-0.07 $\pm$ 0.02 & 2.38	 $\pm$	0.06	&	5.01	 $\pm$	0.09	&	1.36		 $\pm$	0.15	\\
0.9-1.0	&	0.485 $\pm$ 0.005 & 0.45		 $\pm$	0.01	&	0.32		 $\pm$	0.01	&	0.21		 $\pm$	0.01	&	-0.08 $\pm$ 0.01 & 2.33		 $\pm$	0.03	&	4.95		 $\pm$	0.05	&	1.20		 $\pm$	0.07	\\
\hline
\end{tabular}
\end{table*}

Our observed Fourier parameters for SS For clearly fall within the ranges of Fourier
parameters derived from the $V$ light curves of 257 RRab stars published by
Kov\'acs \& Kanbur (1998).  The ranges we obtain also intersect with the theoretical
parameters calculated by Feuchtinger (1999) and Dorfi \& Feuchtinger (1999,
their Figure\,8 and 11). Note that the latter authors express their Fourier parameters in the cosinus
frame.  
For the amplitude ratios we have values in the range $R_{21}=0.41-0.55$,
$R_{31}=0.27-0.34$, $R_{41}=0.15-0.25$ for the $V$ data.
The phase differences in the cosinus frame are in the ranges
$\phi_{21}=3.67-4.06$, $\phi_{31}=1.44-1.99$, $\phi_{41}=-0.82-0.05$.

\begin{figure}
\includegraphics[width=9cm, angle=0]{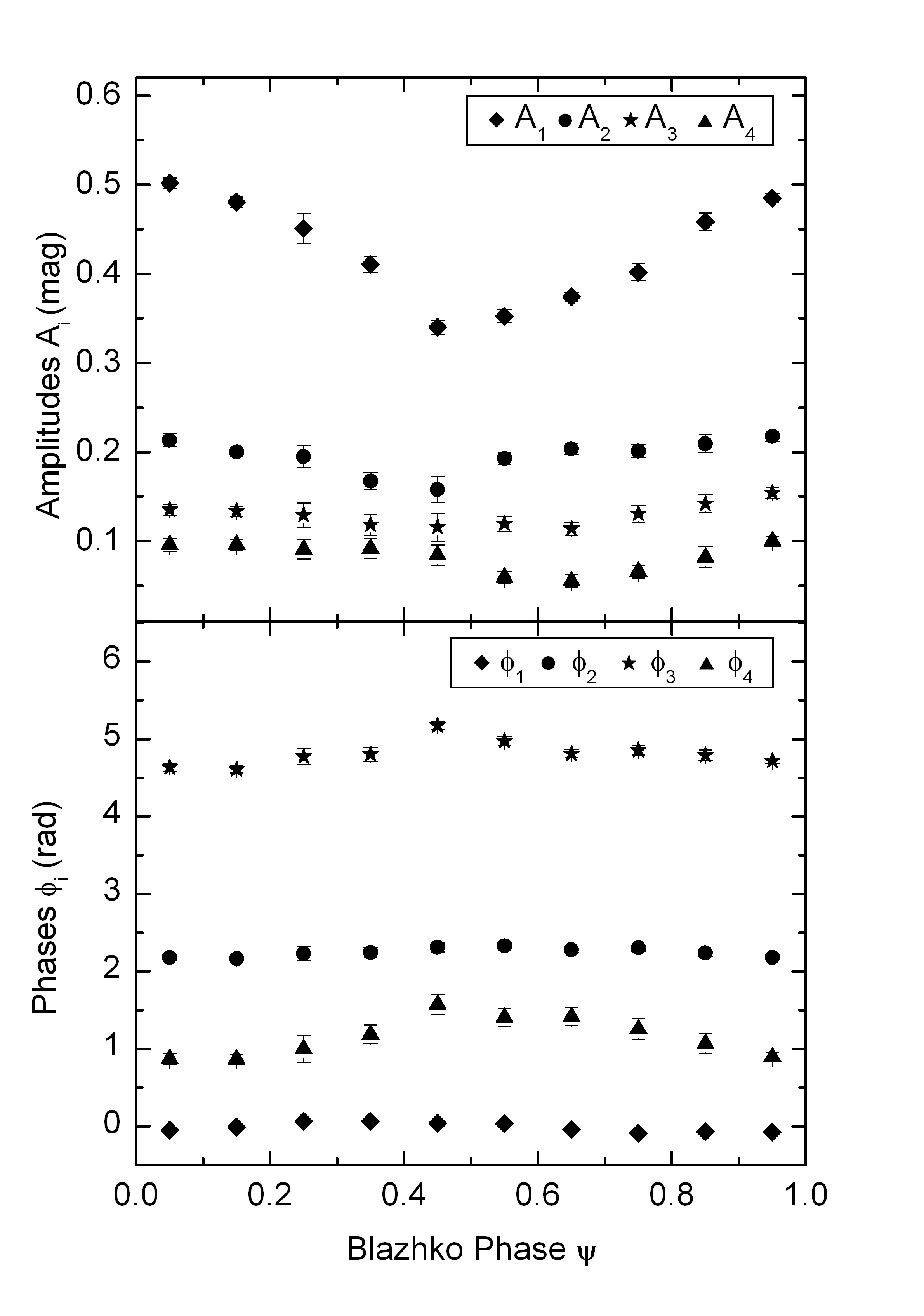}
 \caption{Variations of the amplitudes $A_k$ and the phases
$\phi_k$ over the Blazhko cycle for $k=1,...,4$. }
\end{figure}
\begin{figure}
\includegraphics[width=9cm, angle=0]{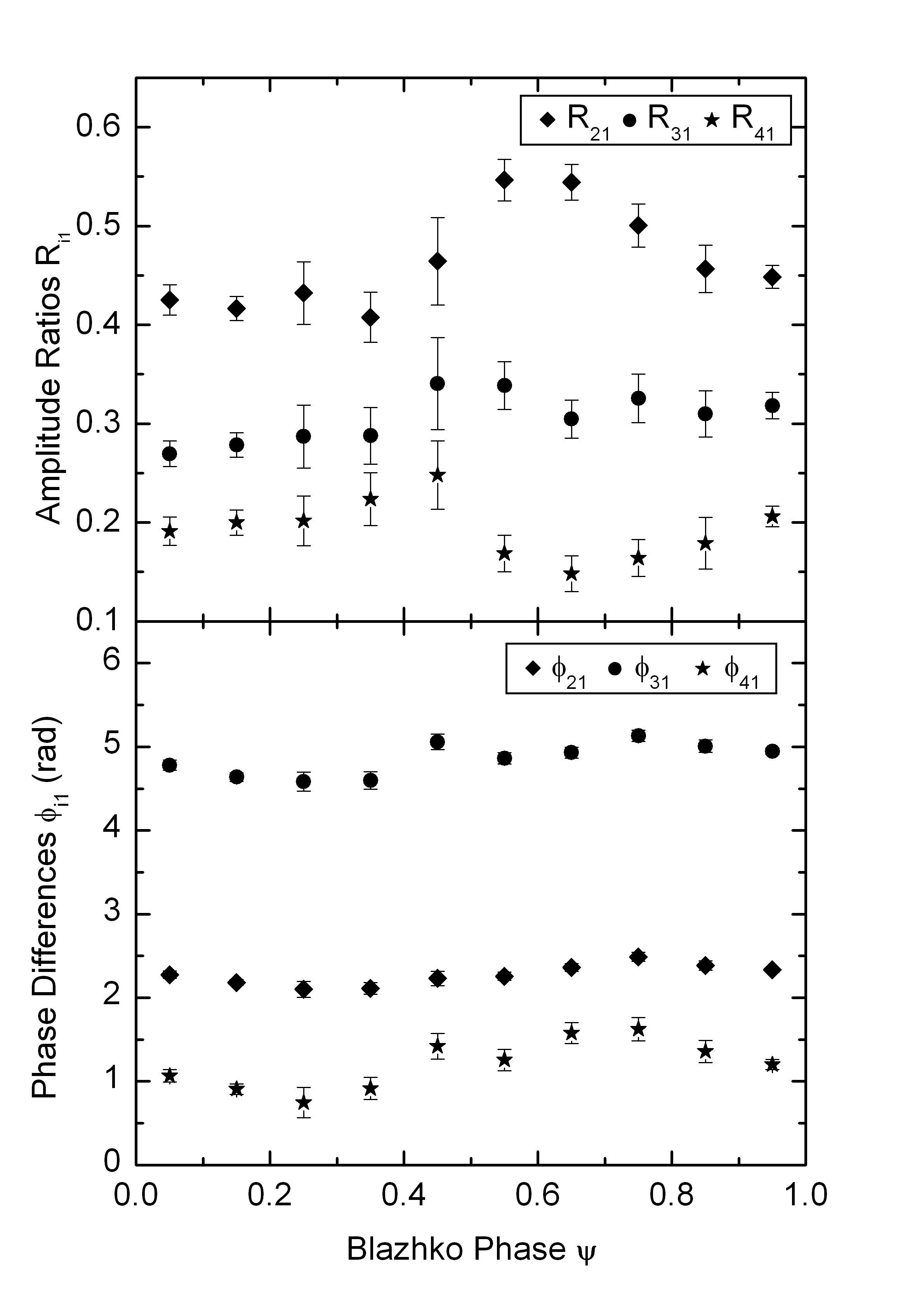}
 \caption{Variations of the Fourier parameters $R_{k1}$ and $\phi_{k1}$ over
 the Blazhko cycle for  $k=1,...,4$. }
\end{figure}

\section{Discussion}

\subsection{Blazhko period}

The Blazhko character of SS For was recognized by Lub (1977), who noticed
strong variations around minimum light of the star.  At that time, its Blazhko 
period was not known.
Wils \& Sodor (2005) published a study on Blazhko variability from the ASAS
database. 
SS For appears in their table with elements of new and suspected Blazhko RRab
stars.  They determined a Blazhko period of 34.8 d.
From our new data set we find a value for the Blazhko period
$P_B=34.94 \pm 0.05$ d, as the difference between $f_0+f_B$ and $f_0$.

\subsection{Multiplets challenge the models}

The observed asymmetry of the modulation components in the {\bf triplets}, also observed in SS
For (see Section\,3.3.1), remains unexplained by both the resonance model and the magnetic model, at least in the degree it is observed.

In our data we find clear evidence for {\bf quintuplet components} in the vicinity of the harmonics of the main frequency in our data.  A quintuplet structure is generally
predicted by the magnetic model (Shibahashi \& Takata 1995), where the magnetic field causes the main radial mode to deform and have additional quadrupole ($\ell = 2$) components.  For more than a decade after the magnetic model was first presented, no evidence for quintuplet components was found from any data set.  The argument given by proponents of the magnetic model was that a quintuplet structure may
manifest itself as only a triplet depending on the geometrical configuration (angles
of pulsation axis, magnetic axis and aspect angle).  Also, the quintuplet components might have such a low amplitude that they are hidden in the noise of the frequency spectrum. This is supported by recent findings by Hurta et al. (2008) in RV UMa, Jurcsik et al. (2008) in MW Lyr, and this work, but it does not imply the magnetic model is the one to be preferred over the resonance model.

A magnetic field of about 1 kG is needed in the magnetic model for the amplitude modulation to be observable 
(Shibahashi \& Takata 1995). The only Blazhko star for which a dedicated spectropolarimetric campaign 
has been carried out so far is RR Lyr, the prototype of the class, and the results are rather contradictory. Babcock (1958) and Romanov et al. 
(1994) reported a variable magnetic field in RR Lyr with a strength up to 1.5 kG, whereas Preston (1967) 
and Chadid et al. (2004) contradict these measurements. The detection of magnetic fields in RR Lyrae 
stars has been hampered by the fact that these stars are quite faint; RR Lyr is by far the brightest 
with V = 7.2-8.2. With the new generation of powerful spectrographs attached to big telescopes, however, many 
more RR Lyrae stars have now come within reach for spectropolarimetry. 


Interestingly, in this data set we only find clear evidence for quintuplet components around the harmonics of the main frequency and not around the main frequency itself. This may be due to the fact that the observed amplitude spectrum is also influenced by the coverage of our measurements (see Jurcsik et al. 2005a).  Satellite data with quasi-uninterruped coverage over several Blazhko cycles will give an answer to many questions concerning the frequency spectrum of Blazhko stars, the modulation components and the occurrence of quintuplet components.

\subsection{Physical parameters from the light curve}

Jurcsik \& Kov\'acs (1996), herafter JK96, determined a $P-\phi_{31}-$[Fe/H] relation for
fundamental mode RR Lyrae stars. 
Their calibration was based on a sample of
81 RRab field Blazhko stars, and its application to RR Lyrae stars
with independent spectroscopic metallicities has proven that it has an overall
prediction accuracy of 0.12 dex (Kov\'acs 2005).  The question, however, is
whether this empirical formula is also applicable to amplitude- and phase-modulated fundamental
mode RR Lyrae stars. 

According to Layden (1994), SS For has a
spectroscopic iron abundance [Fe/H]$_{\rm spec}$= -1.35, which is -1.09 on the scale of the JK96 relation.
Based on ASAS data, Kov\'acs (2005) calculated the iron abundance
[Fe/H]$_{\rm Four}$ according to the JK96 formula, and found a
$\Delta$[Fe/H]=[Fe/H]$_{\rm spec}$-[Fe/H]$_{\rm Four}$ of 0.44.  Kov\'acs
(2005) attributed the large
deviation both to an inaccurate spectroscopic iron abundance
determination, as well as the light curve shape and therefore Fourier parameter 
variations due to the Blazhko effect.
We support the latter explanation, since
from our determination of the Fourier parameters at different phases in the
Blazhko cycle, we obtain values for $\phi_{31}$ between 4.58 and 5.13 rad
(in the sinus frame, as required for the JK96 formula). This leads to
metallicities [Fe/H]$_{\rm Four}$ between -1.55 and -0.81, 
a rather large range but with an average (-1.18) rather close to the spectroscopic value.

Based on the result of Alcock et al. (2003), it is possible that the average
light curves of Blazhko stars can be employed equally well in the empirical
formula calibrated on strictly periodic stars. However, long-term monitoring
- from months to years - is needed to achieve sufficient coverage for Blazhko
variables.  The test can be done for SS For, for which we obtained average light
curve parameters with a high accuracy. For the mean light curve, this yields the value
$\phi_{31}=4.935 \pm 0.057$ (this corresponds to 0.785 $\pm$ 0.009 in
fractions of $2\pi$, as can be derived from Table\,2).  Application of the
JK96 calibration then yields [Fe/H]=-1.07 $\pm$ 0.08, agreeing very well
with the value -1.09 by Layden (1994) on the JK96 scale.

\subsection{Variations at minimum light}


While in most Blazhko stars the variations of the pulsation maximum are
most pronounced, in SS For also the minimum shows strong variations during
the Blazhko cycle. These variations have already been reported by Lub (1977) and
have been investigated in more detail by Guggenberger \& Kolenberg (2006) on
the basis of the data presented in this paper.
They found a periodic behaviour of the so-called bump in the light curve that appears just before minimum
light. The bump occurs at an earlier phase around Blazhko minimum and later
around Blazhko maximum. Also the strength of the bump shows a dependence on
the Blazhko phase: the bump is most distinct during minimum and almost
vanishes during Blazhko maximum.

The bump has been explained by a collision between the free-falling high
atmospheric layers and the deep atmosphere, which has a smaller infall
velocity (Gillet \& Crowe 1988).
From spectroscopic data of RR Lyr, Preston, Smak, \& Paczynski (1965) concluded that a
displacement of the shock forming region occurs over the Blazhko cycle.  This
may induce an earlier or later occurence of the bump in the light curves.
Also the U-B excess observed on the rising branch, as well as the strength of
the H emission (Hardie 1955), both linked to the shocks in RR Lyrae stars, are variable over
the Blazhko cycle (Preston, Smak, \& Paczynski1965).

\section{Conclusions}

High-precision photometric observations in the Johnson $B$ and $V$ filter were obtained from
the South African Astronomical Observatory (SAAO) and Siding Spring
Observatory (SSO), Australia, between October 2004 and September 2005. 
The time span of the data covers almost 10 complete Blazhko cycles.

A detailed analysis of the combined standardized photometric measurements led to the
following conclusions: 

\begin{itemize}

\item  From a Fourier analysis of the combined SAAO+SSO data set we
  obtain $f_0 = 2.018433 \pm 0.000001 $ 
cd$^{\rm -1}$ for the radial pulsation frequency and 
$f_N = 2.04705 \pm 0.00004$ 
cd$^{\rm -1}$ for the right-side peak frequency.  

\vspace{2mm}
\item {\it From our data} we find a value for the Blazhko period 
$P_B=34.94 \pm 0.05$ d.

\vspace{2mm}
\item The amplitudes in the $B$ data are a factor of about 1.2 larger than in the $V$ data.  
The harmonics of the main frequency as well as the triplet 
  structure around the main frequency and its harmonics are
  detectable up to high order, especially in the $B$ data (15th order).  
  Also the Blazhko modulation is a factor of 1.2 larger in the $B$ than in the $V$ data.
  
\vspace{2mm}
\item After subtracting the fit including triplet components, we find clear evidence of quintuplet frequencies in the data.  We found several clearly significant peaks located at $kf_0 \pm 2f_B$ frequency values.  These frequencies emerged most clearly from the $B$ data set in which the modulation components have higher amplitudes, but were also retrieved from the $V$ data. 
Our data thus clearly show the existence of quintuplet components in SS For, as were also found by Hurta et al. (2008) in RV UMa and by Jurcsik et al. (2008) in MW Lyr. 
Quintuplet components are predicted by the magnetic model for explaining the Blazhko effect (Shibahashi \& Takata 1995), but so far there has been no convincing, unambiguous positive detection of a magnetic field in RR Lyrae stars. 

\vspace{2mm}
\item We also find evidence for a so-called septuplet component in our data (at $6f_0+3f_B$).  Multiplet components of order $j \ge 3$ in $kf_0 \pm jf_B$ have also been found by Jurcsik et al. (2008) in their extensive data set of MW Lyr.  They have yet to be explained by the models for the Blazhko effect. Uninterrupted satellite data of Blazhko stars, covering several Blazhko cycles, such as delivered by the COROT satellite (Baglin 2007), will undoubtedly shed new light upon the existence of higher-order multiplet structures in Blazhko stars. We may expect to find many more in high-quality data sets.

\vspace{2mm}  
\item A subdivision of the data into 10 Blazhko phase intervals shows the
  variations of the light curve over the Blazhko cycle.  As all the Blazhko phase
  intervals in $V$ have sufficient coverage, we calculated the Fourier parameters and
  plotted their variations.  Observed values at different phases of the
  Blazhko cycle may be useful for
  confrontation with non-linear convective models such as those developed by
  Feuchtinger (1999). It would be worthwhile to find out whether the observed
  variations in the Fourier parameters can be reproduced theoretically for a
  star with constant mass and metallicity. 
  
 \vspace{2mm} 
\item Application of the empirical $P-\phi_{31}-$[Fe/H] relation developed by
  Jurcsik \& Kov\'acs yields a good result for the average light curve
  derived from all our data. This strengthens the assumption that the average light curve of Blazhko
  stars can be employed in the empirical formula calibrated by Jurcsik \&
  Kov\'acs on stricly periodic stars.  However, the average light curve of a
  Blazhko star can only be determined accurately if the Blazhko cycle is
  sufficiently covered by the available data.

\vspace{2mm}
\item From a study of the light curve shape at different Blazhko phases we
  clearly see that the strong variations around minimum light in SS For are
  related to the position and strength of the bump. A detailed
  study of the bump behaviour over the Blazhko cycle, in photometric as well
  as spectroscopic data, may shed new light upon
  the understanding of the Blazhko effect.

\vspace{2mm}
 Theoretical efforts to revise or expand the existing models for the Blazhko effect would be worthwhile, as well as the exploration of alternative explanations. 

\end{itemize}

\begin{figure*}
\includegraphics[width=12cm, angle=0]{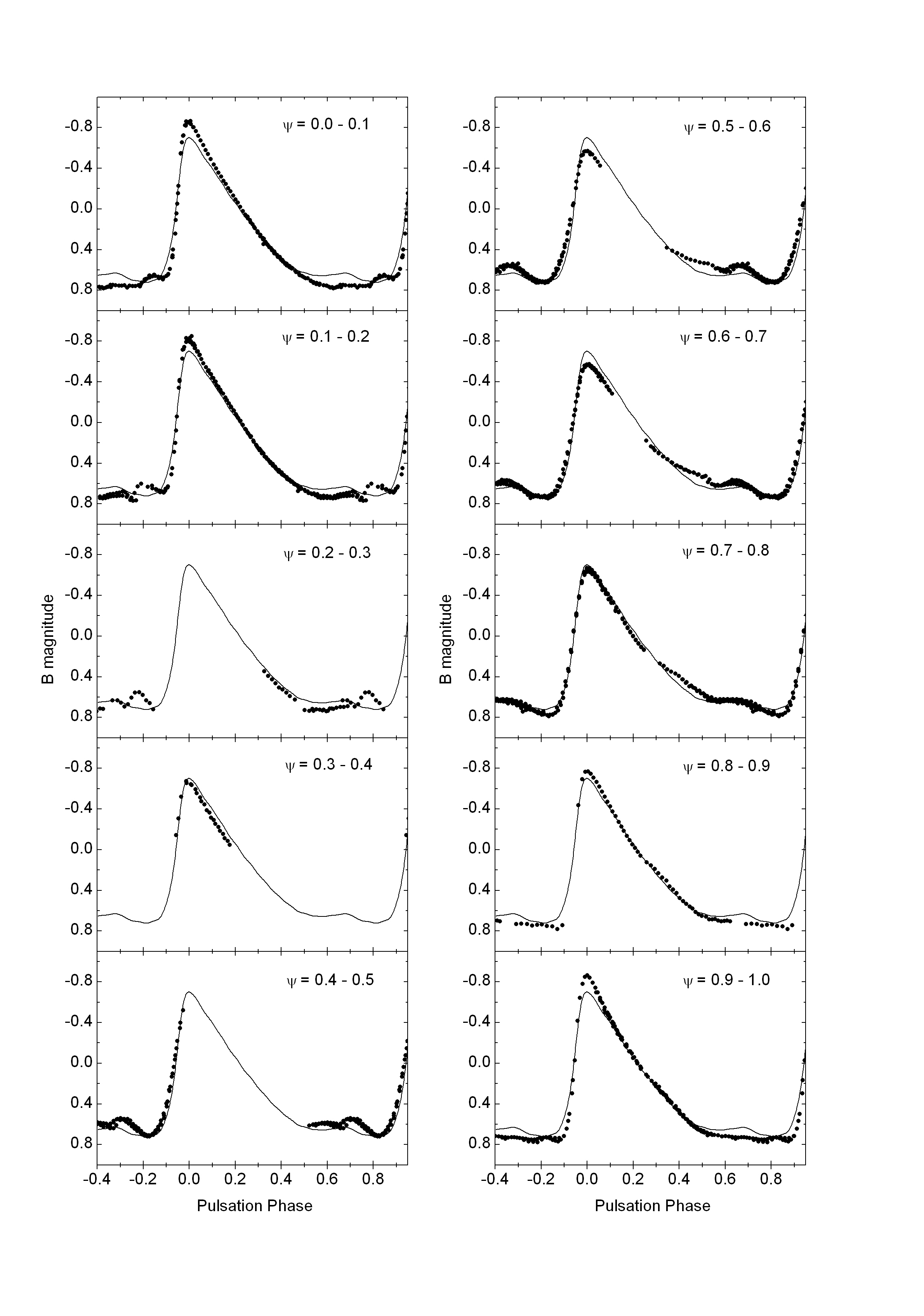}
 \caption{$B$ light data (crosses) in different 0.1 phase intervals of the
 Blazhko cycle (34.94 $\pm$ 0.05 d from this data set).  The mean light curve over all Blazhko phase
 intervals is also shown as a solid line.  Note the strong variations at the
 bump phase before minimum light. }
\end{figure*}

\section*{Acknowledgments}
We thank the referee, J. Jurcsik, for useful suggestions. Part of this investigation has been supported by the Austrian Fonds zur 
F\"orderung der wissenschaftlichen Forschung, project number P19962-N16 and T359-N16 (the Blazhko Project, Kolenberg 2004).
This paper uses observations made at the South African Astronomical
Observatory (SAAO), South Africa, and Siding Spring Observatory (SSO),
Australia.
The research has made use of the SIMBAD astronomical database 
(http://simbad.u-strasbg.fr/), the GEOS RR Lyrae database 
(http://dbrr.ast.obs-mip.fr/), the ASAS database
(http://archive.princeton.edu/\~\,asas/) and the HIPPARCOS catalogue
(http://www.rssd.esa.int/hipparcos/research.html).

\end{document}